\documentclass[fleqn,usenatbib]{mnras}

\usepackage{newtxtext,newtxmath}

\usepackage[T1]{fontenc}

\DeclareRobustCommand{\VAN}[3]{#2}
\let\VANthebibliography\thebibliography
\def\thebibliography{\DeclareRobustCommand{\VAN}[3]{##3}\VANthebibliography}

\usepackage{graphicx}	
\usepackage{amsmath}	

\usepackage{amssymb}	

\def\purple#1 {{\textcolor{purple}{#1}}\ }
\def\red#1 {\textcolor{red}{#1}}
\def\new#1 {{\bf #1 }}
\def\blue#1 {{\textcolor{blue}{#1}}\ }
\def\zy#1 {\textcolor{red}{ zy: #1}}
\def\xycao#1 {\textcolor{red}{ xiaoyue: #1}}
\def\lei#1 {\textcolor{blue}{ lei: #1}}

\newcommand{\angstrom}{\textup{\AA}}

\title[Radio Jet of Cloverleaf]{Discovery of a radio jet in the Cloverleaf Quasar at z = 2.56}

\author[L. Zhang et al.]{
Lei Zhang             ,$^{1,2}$
Zhi-Yu Zhang          ,$^{1,2}$ \thanks{E-mail: zzhang@nju.edu.cn}
James. W. Nightingale ,$^{3}$
Ze-Cheng Zou          ,$^{1,2}$
Xiaoyue Cao           ,$^{4,5,6}$
\and
Chao-Wei Tsai         ,$^{5}$
Chentao Yang          ,$^{7,8}$
Yong Shi              ,$^{1,2}$
Junzhi Wang           ,$^{9,10}$
Dandan Xu             ,$^{11}$
Ling-Rui Lin          ,$^{1,2}$
\and
Jing Zhou             ,$^{1,2}$
Ran Li                ,$^{6,5,4}$
\\
$^{1}$  School of Astronomy and Space Science, Nanjing University, Nanjing 210023, People’s Republic of China\\
$^{2}$  Key Laboratory of Modern Astronomy and Astrophysics (Nanjing University), Ministry of Education, Nanjing 210023, People’s Republic of China \\
$^{3}$   Department of Physics, Institute for Computational Cosmology, Durham University, South Road, Durham DH1 3LE, UK\\
$^{4}$   School of Astronomy and Space Science, University of Chinese Academy of Sciences, Beijing 100049, People’s Republic of China \\
$^{5}$   National Astronomical Observatories, Chinese Academy of Sciences, 20A Datun Road, Chaoyang District, Beijing 100012, People's Republic of China\\
$^{6}$   Institute for Frontiers in Astronomy and Astrophysics, Beijing Normal University, Beijing 102206, People’s Republic of China \\
$^{7}$   Department of Space, Earth and Environment, Chalmers University of Technology, Onsala Space Observatory, SE-439 92 Onsala, Sweden \\
$^{8}$   European Southern Observatory, Alonso de Córdova 3107, Vitacura, Casilla 19001, Santiago de Chile, Chile\\
$^{9}$   Shanghai Astronomical Observatory, Chinese Academy of Science, 80 Nandan Road, Shanghai 200030, People's Republic of China\\
$^{10}$  Guangxi Key Laboratory for Relativistic Astrophysics, Department of Physics, Guangxi University, Nanning 530004, PR China\\
$^{11}$  Department of Astronomy, Tsinghua University, Beijing 100084, China
}

\date{Accepted XXX. Received YYY; in original form ZZZ}

\pubyear{2023}

\begin{document}
\label{firstpage}
\pagerange{\pageref{firstpage}--\pageref{lastpage}}
\maketitle

\begin{abstract}
The fast growth of supermassive black holes and their feedback to the host
galaxies play an important role in regulating the evolution of galaxies,
especially in the early Universe. However, due to cosmological dimming and the
limited angular resolution of most observations, it is difficult to resolve the
feedback from the active galactic nuclei (AGNs) to their host galaxies.
Gravitational lensing, for its magnification, provides a powerful tool to
spatially differentiate emission originating from AGN and host galaxy at high
redshifts.  Here we report a discovery of a jet-like radio structure in a strongly lensed
starburst quasar, H1413+117 or Cloverleaf at redshift $z= 2.56$, based on
observational data at optical, sub-millimetre, and radio wavelengths. With both
parametric and non-parametric lens models and with reconstructed images in the
source plane, we find a well-separated, kpc-scaled, single-sided radio jet located at projected ${\sim}1.2\,\mathrm{kpc}$ to the northwest of the host galaxy
in the source plane. 
This could indicate the co-existence of feedback from the AGN by both wind and jet in the Cloverleaf quasar.
\end{abstract}




\section{Introduction}

The correlated growth of the bulge mass of a galaxy and its supermassive black
hole (SMBH) mass throughout cosmic time has been described as their
co-evolution \citep{Ferrarese2005, Kormendy2013}.  When the SMBH actively
accretes, the active galactic nucleus (AGN) turns on, which can drive powerful
feedback to the surrounding interstellar medium (ISM) of the host galaxy.  The
radiation pressure and the energetic particles in the AGN wind and jet behave
as the main forms of feedback. The feedback process can sweep and heat up
the cold gas along its radial paths and, as a result, quench the star formation
\citep{Springe2005, McNamara2007, Cattaneo2009}.  For the high-luminosity
quasars, the host galaxy and surrounding environment receive powerful feedback
from its AGN through electromagnetic radiation (known as quasar mode), kinetic
jets (known as radio mode), or both \citep{Fabian2012, Kormendy2013,
Churazov2005}.

Similar to the cosmic star formation, the evolution of quasi-stellar object
(QSO) activity also shows a strong cosmic evolution, reaching its peak at
$z\sim2$ \citep[e.g.,][]{Schmidt1983, Wolf2003}. The feedback of QSOs in this
era is therefore considered as the key mechanism to quench star formation.
Galactic outflows driven by AGNs \citep[e.g.,][]{Farrah2012, Cano2012,
Nesvadba2011, Alexander2010, Prochaska2009, Nesvadba2008} have shown evidence
of quenching. High-velocity galactic winds, which are commonly detected with
the broad absorption lines against a strong quasar continuum
\citep[e.g.,][]{Dunn2010, Moe2009, Saez2009} and also evidenced with
highly-ionized Fe lines in low-luminosity AGNs \citep{Shi2021}, can efficiently
remove the ISM from the host galaxies. On the other hand, radio lobes/jets
\citep[e.g.,][]{Gopal-Krishna2001, McNamara2007, Cattaneo2009MN, Cattaneo2009,
Fabian2012}, which are often discovered on scales from several tens of kpc to
Mpc, can also directly impact star formation.

However, the majority of high-redshift QSOs have very compact sizes
\citep{Silverman2019, Li2021, Ding2022}, which limits the separation between
their host galaxy and AGN-related components. Multiple-wavelength observations
with spatial resolutions on scales of sub-kpc are therefore needed, which are
still extremely difficult, given the capability of current instruments and
cosmological dimming effects \citep{Lanzetta2002}.

Gravitational lensing serves as a powerful tool to improve both sensitivity and
spatial resolution. With proper demagnification modelling, the lensing system
allows us to resolve the detailed galactic structures of high-$z$ QSOs for case
studies.  Cloverleaf (also known as H1413+117) is one such target. Discovered
by \citet{Magain1988}, it is a strongly lensed starburst galaxy with a bright
AGN at $z = 2.558$. It contains four images at both optical and sub-millimetre
wavelengths \citep{Kneib1998}. With the detections of CO, HCN, and FIR
emission, Cloverleaf is confirmed to contain an extensive molecular disc of
${\sim}1.6$\,kpc in diameter, a molecular gas mass of $\sim 10^{10}\,\rm
M_{\odot}$, and a star formation rate of $10^3\,\rm M_{\odot}\,yr ^{-1}$
\citep{Solomon2003, Venturini2003, Barvainis1997, Weiss2003}. On the other
hand, information about this system in the lens plane is limited due to the
lens galaxy's faint emission \citep{Lawrence1996, Angonin1990, Turnshek1997},
while \citet{Chantry2007} suggested a deconvolved lens image from Hubble Space
Telescope (HST) {\it{F160W}} and {\it{F160W}} data. The redshift of the lens
galaxy is derived from the H{\sc i} narrow absorption systems
\citep{Turnshek1988, Magain1988} at $z \sim 1.7$.

With the latest high-sensitivity, high-resolution, and high-fidelity data of
Cloverleaf system from ALMA, VLA, and HST, we are able to not only compare
morphologies from different gas phases but also constrain the lens galaxy mass
models more accurately than previous models. The traditional method
\citep[e.g.][]{Kneib1998}, which is based on parametric modelling, has limitations in
the cases of irregular sources. The newly developed non-parametric modelling,
which is required for high angular resolution and high sensitivity analysis,
greatly improves the reconstructed source plane \citep{Nightingale2015,
Nightingale2018} and thus, the structures of the Cloverleaf system can be well
distinguished. 

This work focuses on observational data at three wavelengths: (i) optical,
which is dominated by the QSO emission and is seen as almost four point sources
from HST images \citep{Kneib1998}; (ii) Sub-millimetre, which traces the
gaseous dusty disk of the host galaxy \citep{Solomon2003, Granato1996}, with
current ALMA observations having enough resolution to resolve the four images
and; (iii) radio, where the radio emission
traces relativistic charged particles accelerated by the radio jet
\citep{Kayser1990} launched by the central AGN or starburst.  
With these data, we can separate the AGN contribution from the host galaxy.

This paper is organized as follows: Observations and data reduction are
presented in Section~\ref{sec:obsdata}. The modelling processes, the lens
models of the host galaxy, and the newly discovered radio jet, which are built
up with both parametric and non-parametric models, are shown in
Section~\ref{sec:model}. In Section~\ref{sec:result}, we analyse the spatial
distribution of multi-wavelength images and the spectral energy distribution (SED) of radio emission. In Section~\ref{sec:discuss}, we discuss the
confirmation, emission components, and feedback of this radio jet. Last, we
conclude the summary in Section~\ref{sec:conclusion}. 

Throughout the paper, we assume a cosmology with $\Omega_\mathrm{m} = 0.310$,
$\Omega_{\Lambda} = 0.689$, and $H_0 = 67.7\,\mathrm{km\,s^{-1}\,Mpc^{-1}}$
\citep{Planck2020}.  In this case, $1''$ corresponds to $\sim$8.7\,kpc in the
lens plane at $z = 1.7$ and $\sim$8.22\,kpc in the source plane at $z = 2.56$. 

\section{Multi-band Observational Data and Data Reduction} \label{sec:obsdata}

We obtain multi-wavelength, high-angular-resolution data from the archival systems of HST
(optical), ALMA (sub-millimetre), VLA (radio), and e-Merlin (radio).

\subsection{ HST F814W Image}

The optical image is obtained from the HST archive.  This three-orbit
observation was performed on 26th Jan 2000, with the Advanced Camera for
Surveys (ACS) and the F814W filter. We choose the ACS images for their high
spatial sampling rate of 0.05$''$\,pix$^{-1}$
\citep[][]{Lucas2021acsd.book...10L}. We stacked all ACS images with SWARP
\citep{Bertin2002}. We use {\sc imfit} in Common Astronomy Software
Applications version 6.1.2.7 (CASA) \citep{McMullin2007} to fit a star located
at ($14:15:47.39856, +11:29:50.7480$) (2MASS id: J14154739+1129507
\citep{Cutri2003}) in the field and obtain its full width at half maximum
(FWHM) to be $\sim$0.08 arcsec, which is adopted as the image resolution. The
rms noise is evaluated to 0.2 $\mathrm{erg\,s^{-1}\,cm^{-2}\,{\angstrom}^{-1}}$ (below
referred as counts). 

\subsection{ ALMA 300 GHz continuum}

We adopt the ALMA archival data from projects 2012.1.00175.S (PI: van der Werf,
Paul) and 2017.1.00963.S (PI: Sharon, Chelsea), both observed at Band 7 (see
Table~\ref{tab:ALMA_continuum} for observational details).

\begin{table*}
\caption{Details of the observations on Cloverleaf from the ALMA data archive.}
\label{tab:ALMA_continuum}
  \begin{tabular}{lccccccc}
     \hline
     Proj. ID       & Date          & \multicolumn{2}{c}{Calibrators} & \multicolumn{2}{c}{Freq. Coverage} & Baseline       & On-source Time\\
                    &               & Gain          & Flux/BP        & LSB (GHz)      & USB (GHz)          & $k\lambda$ & min \\
     \hline
     2012.1.00175.S & 25 Jun.\ 2015 & Titan         & J1415+1320      & 325.86--329.28 & 338.43--341.46    & 28--1433   & 68 \\
     2012.1.00175.S & 30 Jun.\ 2015 & J1550+054     & J1415+1320      & 276.68--279.62 & 289.50--292.43    & 36--1530   & 9 \\
     2012.1.00175.S & 27 Sep.\ 2015 & J1550+054     & J1415+1320      & 311.93--314.43 & 325.86--327.85    & 40--2414   & 17 \\
     2017.1.00963.S & 22 May.\ 2018 & J1337+1257    & J1347+1217      & 341.42--345.04 & 353.42--357.04    & 14--372    & 5  \\
     \hline
\end{tabular}
\end{table*}

\begin{figure*}
    \centering
    \includegraphics[scale=0.35]{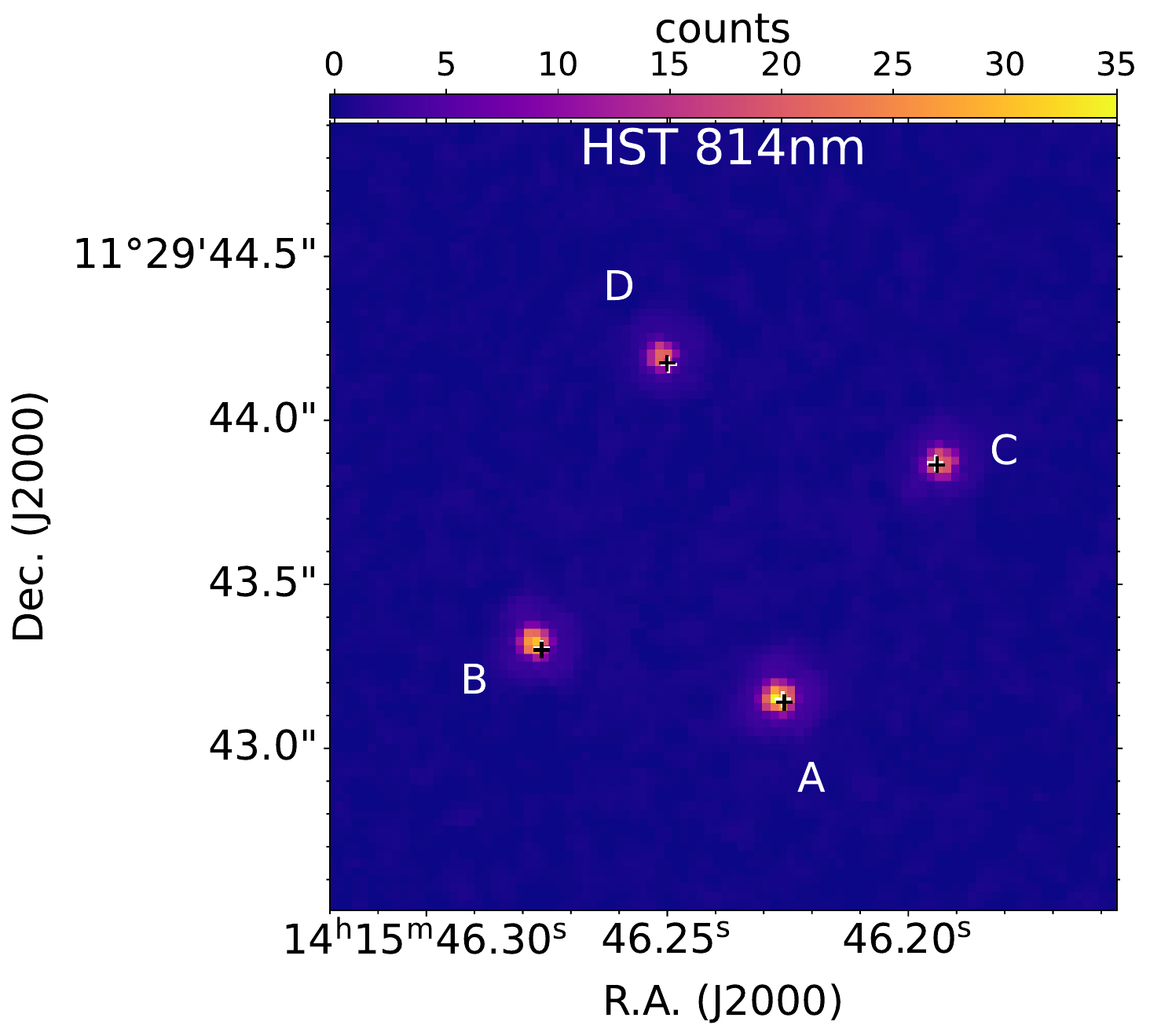}\hfill
    \includegraphics[scale=0.35]{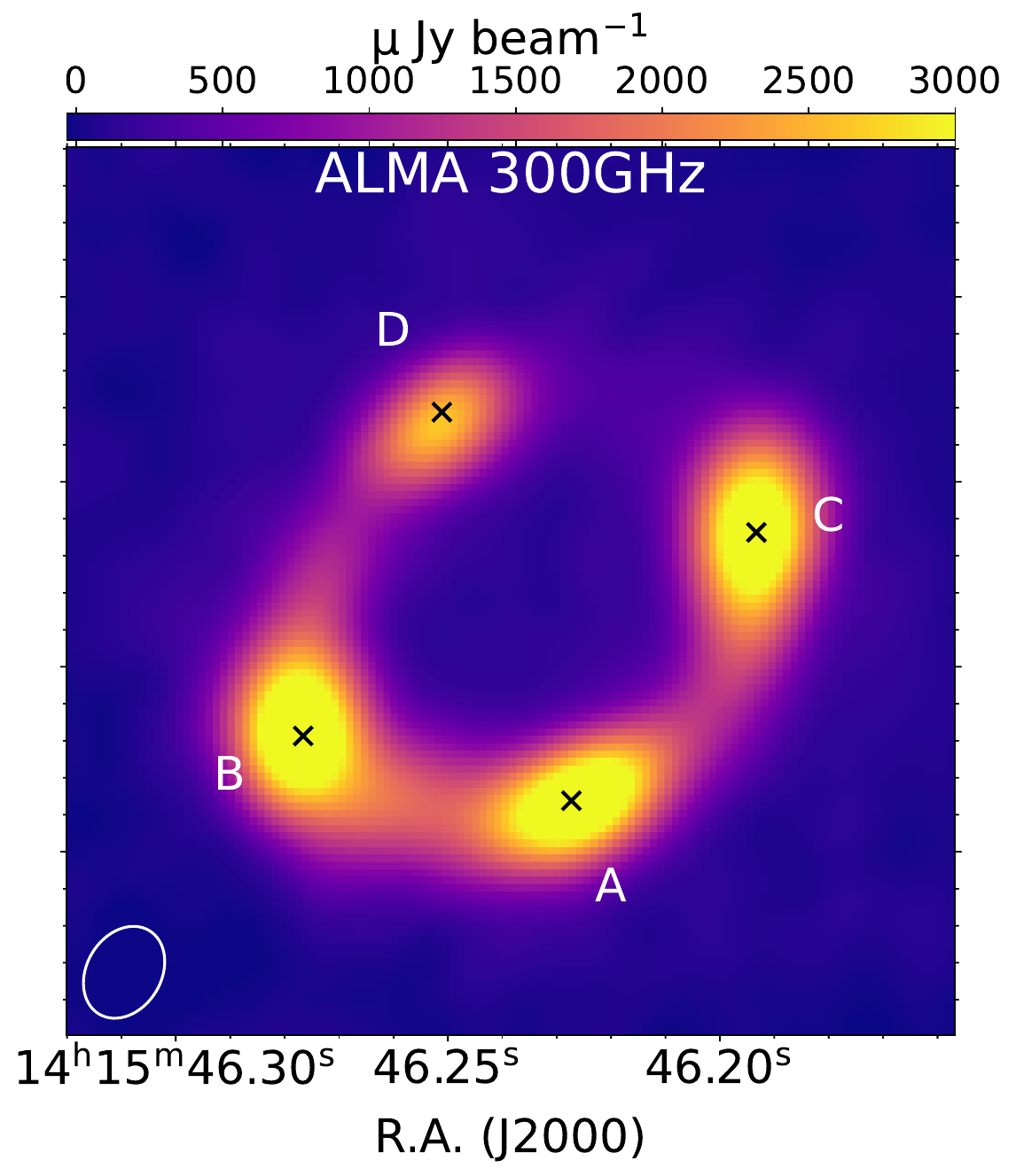}\hfill
    \includegraphics[scale=0.28]{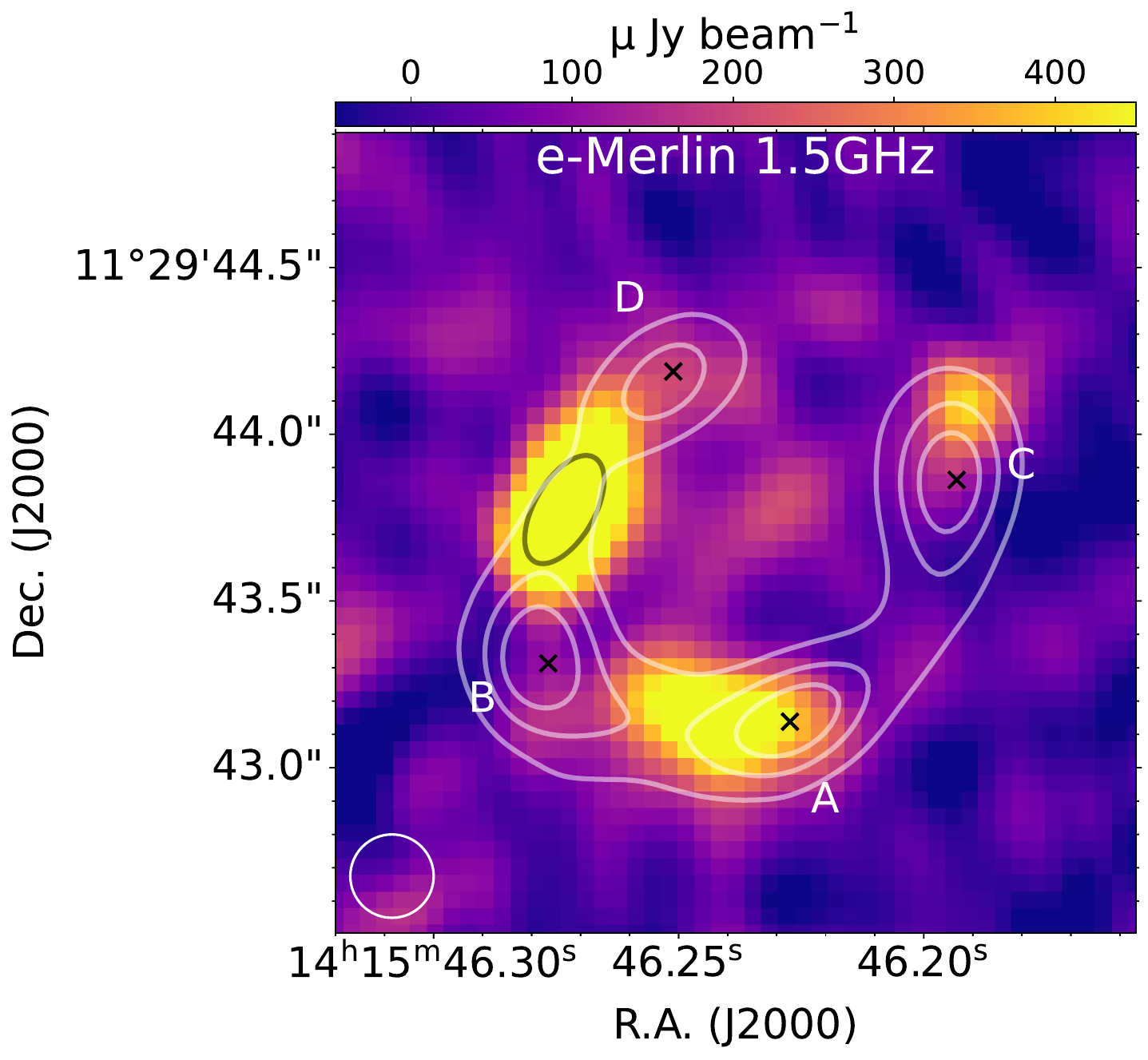}
    \includegraphics[scale=0.28]{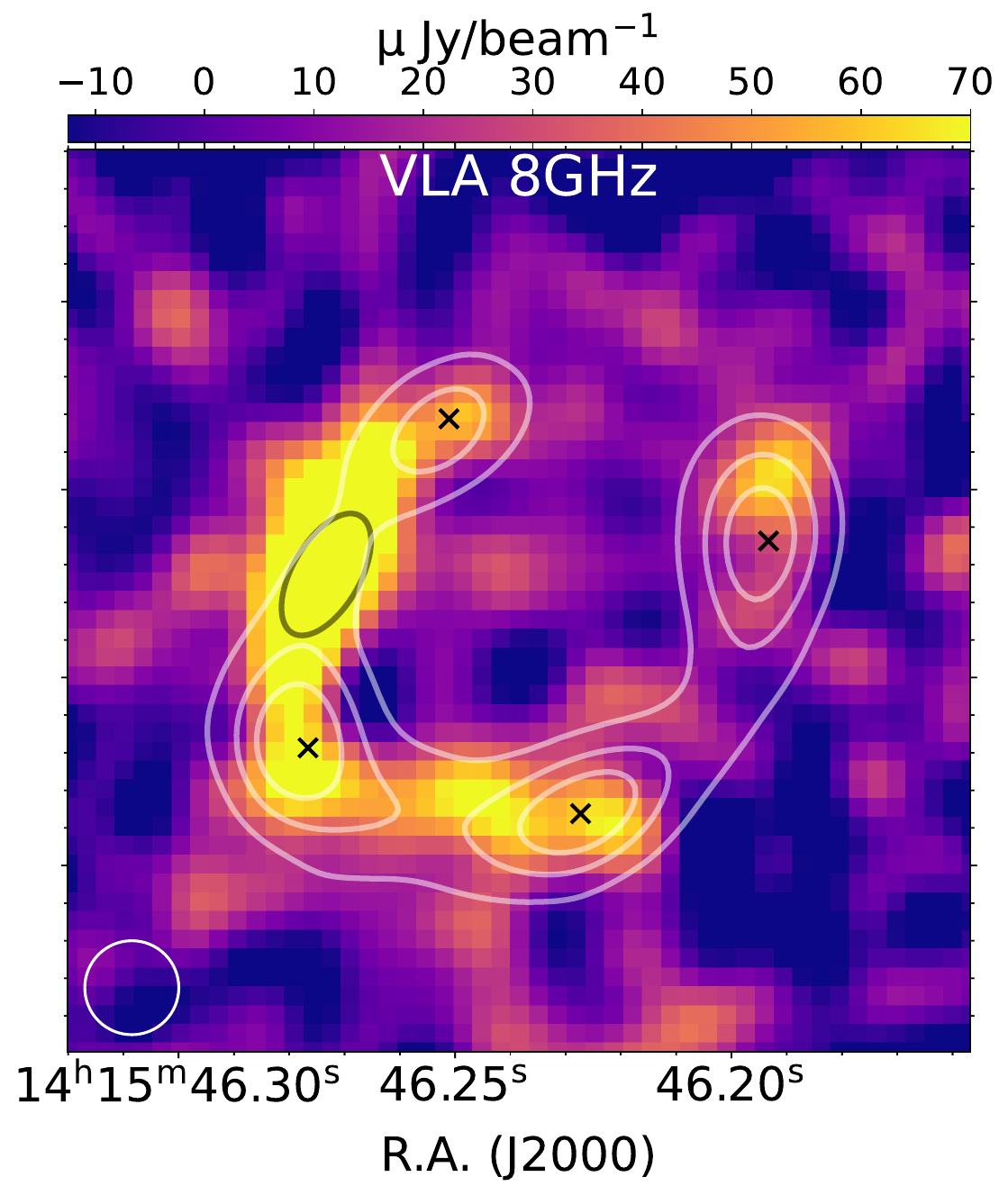} 
    \includegraphics[scale=0.28]{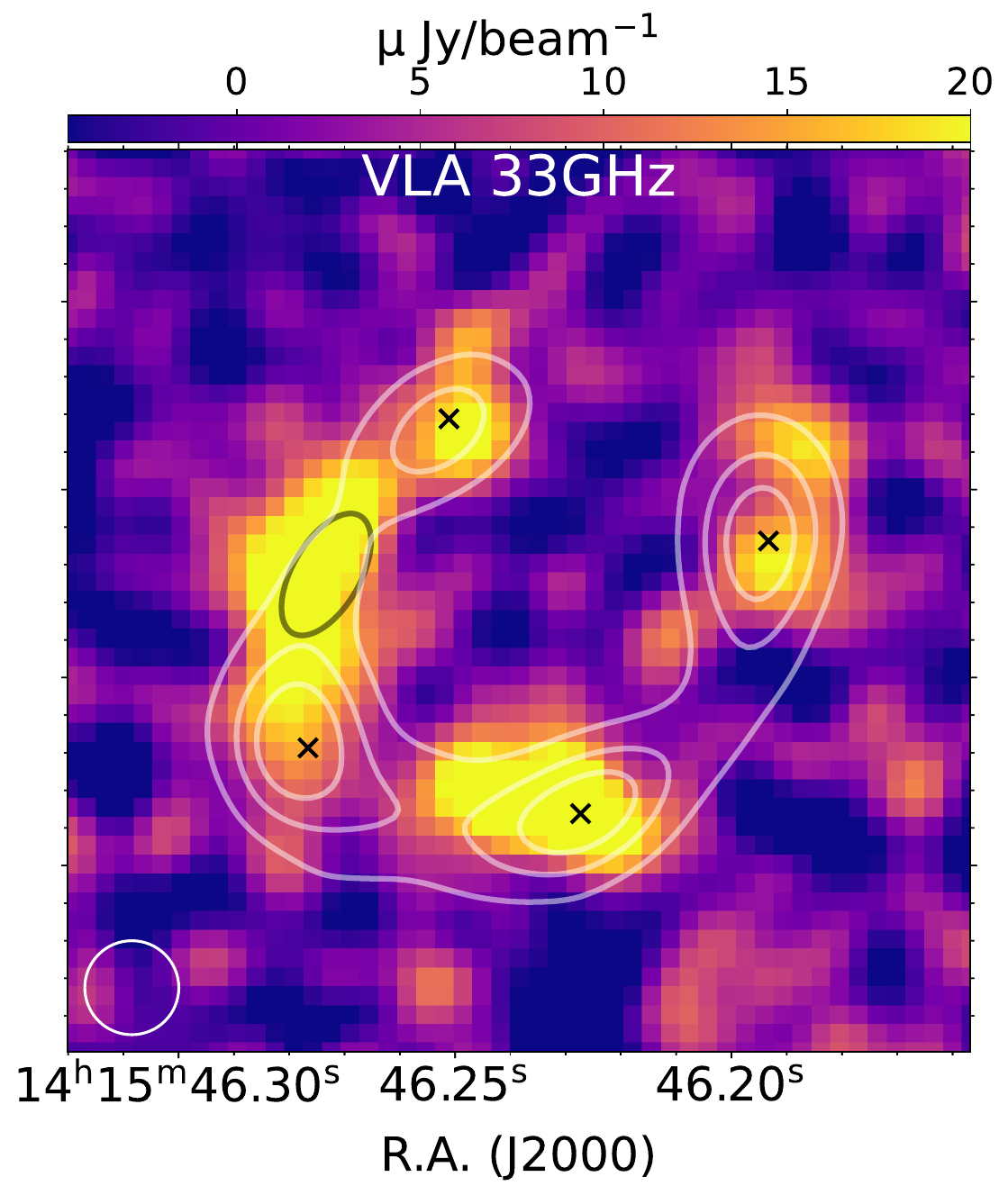}
    \caption{Images of Cloverleaf at optical, sub-millimetre, and radio bands: Top Left: HST optical
    data where components A, B, C and D are four images of this lensing system;
    the black pluses and the white pluses mark the peak positions of the
    model images with the parametric and the non-parametric lens model
    respectively. Top Right: dust continuum data around 300\,GHz where the black
    crosses mark the peak positions of four components in the HST data. 
    Bottom: Radio images of Cloverleaf at 1.5\,GHz (left), 8.4\,GHz (middle),
    and 33\,GHz (right), respectively. Black crosses show the peak
    positions of four components in the HST data. White lines are the contours
    of the ALMA dust continuum, starting from $25\sigma$,
    with steps of $30\sigma$. The black ellipse presents the region 
    we extract fluxes of the radio jet. 
    We use the same restoring beams of $0.25\arcsec$ in all radio images to generate the same source planes for spectral index analysis. }
    \label{fig:dataimage}
\end{figure*}

We calibrate the raw data with CASA \citep[version 6.1.2.7 ][]{McMullin2007}
with the associated standard pipeline. Then we combine all the calibrated
on-source data and omit the frequency range of CO $J$=11-10. We invert the
combined visibility data by task {\textsc{tclean}} with the `mfs' mode, a
gridder of {\textsc{mosaic}}, a cell size of 0.02$''$, and a Briggs weighting
with a {\textsc{robust}} of 1.5. The synthesised image (see
Fig.~\ref{fig:dataimage} top right) has a beam size of $0.26''\times 0.20''$ and a
position angle of $-31\deg$. The noise level is $35\,\rm{\mu Jy\,beam^{-1}}$,
estimated from the emission-free area in the field.

\subsection{ VLA  Archival data} 

\subsubsection{ VLA 8.4 GHz continuum}

The Very Large Array (VLA) 8.4 GHz  data were obtained from the NRAO
archive\footnote{\url{https://data.nrao.edu/portal}}. This data have been originally presented by \citet{Kayser1990}. The project IDs are AS357
and AC243, which were performed on 13 Jan.\ 1989 and 04 Feb.\ 1989,
respectively. Both observations were done with the A-configuration of the VLA.
The total observing time was $\sim$5 hours and 4 hours, with $\sim$1.5 hours
and 1 hour effective on-source time, respectively. Each on-source scan took 15
minutes and 10 minutes with a sampling time of 1 second. In both observations,
3C~286 (1328+307) was adopted as both flux and bandpass calibrators, while the
phase calibrator was 1413+135  in B1950 (or, J1415+1320 in J2000). 
The receivers were configured with two spectral
line windows (SPW), centring at 8.415 and 8.464\,GHz, respectively. Each SPW
covers a 50 MHz bandwidth, offering a total bandwidth of 100\,MHz, with full
stock polarisations. The baselines of both observations range from
15\,k$\lambda$ to 950\,k$\lambda$ in the UV space. 

The data was first imported to {\sc AIPS} and converted to the {\sc uvfits}
format. We use {\sc casa} (ver.\ 6.1.0) to calibrate the data manually
following standard procedures.  We then image the data with task {\sc tclean}
in {\sc casa}, with the `mfs' mode. 
We use a Briggs weighting with a {\sc robust} of 0.5.
The final synthesised beam is $0.25\arcsec
\times0.23\arcsec$ with a position angle of $-47\deg$. We then convert the
equatorial coordinate from the B1950 epoch to the J2000 epoch with task {\sc
imregrid}. The rms noise level is ${\sim}14\,\rm{\mu Jy\,beam^{-1}}$.

\subsubsection{ VLA  33 GHz continuum}

We obtained the Ka band JVLA data from the NRAO archival system. The project
number is 13B-051 (PI: Sharon, Chelsea). This project consists of a total of 18
execution blocks (EBs) targeting Cloverleaf. However, three of them, taken
before 22 Oct. 2013, were corrupted and not usable due to incomplete execution.
The remaining 15 EBs were taken from 27 Oct. 2013 to 5 Jan. 2014, with the
B-array configuration. The pointing observations were performed at the X-band,
and the observations were performed at the Ka band.  These observations
adopted 27 antennas and took a total of $\sim$6.8 hours on-source time.

The receiver covers a frequency range of 31.95--34.00\,GHz, with 16 spectral
line windows (128\,MHz each) and two polarisations. The total bandwidth is
$\sim$2 GHz. The bandpass/flux calibrator and the gain calibrator are 3C~286
and J1415+1320, respectively. The baselines range from 16\,k$\lambda$ to
1192\,k$\lambda$.

We calibrate the raw data using CASA (ver.\ 5.6.2-2) with the standard VLA data
reduction pipeline. Then we combine all calibrated data and abandon the
spectral windows that contain the CO $J$=1$\rightarrow$0 line emission. To save
computational resources, we bin all channels for each spectral line window.
Considering a maximum offset of ${\sim}2''$ from the phase centre, the
bandwidth smearing effect is negligible ($(\Delta \nu / \nu_0) \times
(\theta_{\rm offset} /\theta_{\rm HPBW})  \le 0.026$) for Cloverleaf. Then we
image the visibility data with CASA (ver.\ 6.1.2.7), using task {\sc tclean}.
We adopted the `mfs' mode, a gridder of {\sc standard}, a cell size of
0.05$''$, and a Briggs weighting with a {\sc robust} of 1.5. The cleaned image
has a beam size of $0.26\arcsec \times0.23\arcsec$ with a position
angle of $-37\,\deg$. The rms noise level is ${\sim}4.6\,\rm{\mu Jy\,beam^{-1}}$.

\subsection{E-Merlin 1.5 GHz continuum data}

The e-Merlin 1.5 GHz data was obtained from the e-Merlin archival system. The
project number is CY4215. The observation was performed on 09 Feb.\ 2017, with
a total observing time of 13 hours. The receiver covers a frequency range of
1.25--1.77\,GHz, with eight spectral windows, of which each has 64\,MHz sampled
with 128 channels. The total bandwidth is 512\,MHz. The average sampling time
is 2.0 seconds. 

In total, seven antennas were adopted during the observations. The bandpass and
flux calibrator was 1331+305 (3C286), and the gain calibrator was J1415+1320.
Every 10 minutes, the telescope switch between the target (7 minutes) and the
gain calibrator (3 minutes). The observations spanned $\sim$7.7 hours on source
in total. The baselines range from 24\,k$\lambda$ to 1276\,k$\lambda$ in the UV
space.

The archival data has been correlated but not calibrated. We first convert the
common uvfits format to the format of `measurement set' and use {\sc casa}
(ver.\  6.1.0) to flag the data and calibrate manually. For imaging, we adopted
the `mfs' mode, a gridder of {\sc standard}, a cell size of 0.05 arcsec, and a
Briggs weighting scheme with a {\sc robust} of 1.5. The cleaned image has a
default beam size of $0.26\arcsec \times0.18\arcsec$ with a position
angle of $25\,\deg$. The rms noise level is ${\sim}47\,\rm{\mu Jy\,beam^{-1}}$. 

We also clean the e-Merlin data with a
{\it{nterm}} of 2 to get the spectral index map in the lens plane (see
Fig.~\ref{fig:localspectralindex}).

In the spectral slope index analysis, we use a uniform restoring beam of $0.25\arcsec \times0.25\arcsec$
for all radio images, to analyse these data with the same
reconstructed source plane.  


\subsection{Astrometric correction} 

We notice a systematic astrometric position offset between the HST image and
those at radio/sub-mm bands. 
We fit the peak positions of the four images with SAOimageDS9
\citep{Joye2003} for both the HST data and the ALMA continuum data.  From
these, we get an average offset of 0.126$''$ in the right ascension (R.A.) and
0.219$''$  in the declination (Dec.). After we corrected the offset, the peak
positions of the images obtained from HST and ALMA (see
Fig.~\ref{fig:dataimage}) are still not yet perfectly aligned (though the
current offsets are within one pixel of the HST image), which may originate
from different emission areas and/or imperfections of the observations. 

The radio band observations use the same phase calibrator J1415+1320. 
Their coordinates are almost identical to each other.
The largest offset is lesser than 13 mas, which happens between the VLA 8.4\,GHz and 33\,GHz data. Since the offset is smaller than the pixel scale, we neglect it in the lens modelling below. 

\section{Lens modelling} \label{sec:model}

To analyse the detailed morphology of multi-wavelength images in the source
plane, we adopt the lens model software \texttt{PyAutoLens} (AutoLens)
\footnote{\url{https://github.com/Jammy2211/PyAutoLens}} \citep{pyautolens,
Nightingale2015, Nightingale2018}. We model Cloverleaf with both parametric and
non-parametric models for the lensed source galaxy, following the visual
step-by-step guide provided in  \texttt{PyAutoLens} \footnote{Both parametric
and non-parametric fitting methods are provided as Jupyter notebooks at the
following
link:\url{https://github.com/Jammy2211/autolens_likelihood_function}}.

We take the assumption from \citet{Kneib1998} that the lens system
is contributed by a single galaxy (the narrow absorption systems) and a distant
cluster both at $z \sim 1.7$.

\subsection{Overview of Lens Modelling Strategy}

We perform lens modelling with the following series of linked heuristic steps: 

\begin{itemize}
    \item Fit the HST data using a simple lens model which assumes point-source
            emission for the source galaxy. This provides an initial estimate
            for the parametric lens mass model. 
    \item  Take the highest likelihood results from the previous step as the
            initial input to fit the ALMA 300 GHz continuum data with another
            parametric mass model, using an extended source model which follows
            a smoothly parameterized form of elliptical S\'{e}rsic profile. 
    \item Using the mass model inferred in the previous step to initialise the
            fit, fit different waveband datasets with a non-parametric source
            reconstruction that uses a Voronoi mesh.
\end{itemize}

In our lens model, the redshift of the lens plane is set as 1.7 \citep{Kneib1998}. 

\subsection{Parametric fitting---optical and sub-mm images}

\subsubsection{The HST optical data}

The HST optical (814\,nm) image is shown in the top left panel of
Fig.~\ref{fig:dataimage}.  We use {\sc imfit} to measure the sizes (FWHM) of the
four components (see Table \ref{tab:images table}), by assuming that they
follow Gaussian distributions. The fitted FWHMs of all four components are
almost identical to the FWHM of a point source (${\sim}0.08\arcsec$), so all
these images can be treated as compact point sources without extended
structures, which is also consistent with the result that the optical emission
is dominated by a point-like AGN in the literature \citep{Magain1988}.

First, to model the HST data, we adopt an isothermal ellipsoid with an external
shear for the lens galaxy's mass and use a ``point'' to represent the source
galaxy.  With this optical-band lens modelling, we can preliminary estimate the
lens mass model, as well as the source's flux and position.

The four optical images of Cloverleaf point sources may suffer from
micro-lensing \citep{Kayser1990}, which may lead to a biased result since the
micro-lensing effect is not included in the strong lens modelling.
Additionally, the limited number of data (i.e., flux and position) provided by
the four lensed images only barely constrain the lens model, resulting in
considerable uncertainties. These two limitations beg for a refined analysis at
the submillimeter band, on which we perform the lens modelling with an extended
lensed arc and use the modelling result of the optical band as a prior.  The
extended lensed arc at the submillimeter wavelength is not subject to the
micro-lensing effect and its entire brightness distribution can be used to
improve the precision of the lens model. 

\begin{table*}
    \centering
    \begin{tabular}{lccccc}
    \hline
    Band & ID & Size ($\rm mas\,\times\,mas$) & Peak S/N \\
    \hline
    814 nm & A & 89.4 $\times$ 85.6 & 173\\
    814 nm & B & 84.3 $\times$ 77.5 & 151\\
    814 nm & C & 91.3 $\times$ 86.7 & 106\\
    814 nm & D & 82.4 $\times$ 76.6 & 106\\
    \hline
    300 GHz & A & 540 $\times$ 183 & 116\\
    300 GHz & B & 490 $\times$ 254 & 122\\
    300 GHz & C & 478 $\times$ 216 & 112\\
    300 GHz & D & 426 $\times$ 180 & 75.8\\
    \hline
    1.5 GHz & A & 540 $\times$ 257 & 17.4\\
    1.5 GHz & B\&D & 443 $\times$ 121 & 28.4\\
    1.5 GHz & C & 262 $\times$ 161 & 10.8\\
    \hline
    \end{tabular}
    \caption{Apparent sizes of four images of Cloverleaf at different
    wavelengths, fitted with Gaussian profiles. The sizes are described as (the
    FWHM of the major axis) $\times$ (the FWHM of the minor axis). }
    \label{tab:images table}
\end{table*}

\subsubsection{The ALMA 300 GHz continuum }\label{para_models} 

We next fit the ALMA 300 GHz sub-millimetre data with a more complex parametric
lens model.  We adopt a power-law ellipsoid \citep{Tessore2015} with an
external shear as the lens galaxy mass profile and model the source with an
elliptical S\'{e}rsic profile. 

We fix the centre of the lens mass model (from the result of the HST data
modelling) to reduce the dimensions of parametric space. All remaining
parameters from the previous model are then used to create the initial input
for this modelling. The inferred parameters of the best-fit model at
$1-\sigma$ confidence are shown in Table \ref{tab:parameters of lens},
including the centre positions ($x$ \& $y$), the elliptical components ($e_{\rm x}$
\& $e_{\rm y}$), the Einstein radius ($r$), the slope of the power-law profile
($\alpha$), and the elliptical components of the external shear ($\gamma_{\rm x}$ \&
$\gamma_{\rm y}$).

\begin{table*}
    \centering
    \begin{tabular}{cccc}
    \hline
    Parameters       & Definition                              & Parametric Model                           & Non-parametric Model \\
    \hline
    $x$              & R.A.\ (hh:mm:ss) at mass centre         & 14:15:46.2384 $^{+\ <0.0001}_{-\ <0.0001}$ & 14:15:46.2325 $^{+\ <0.0001}_{-0.0002}$\\
    $y$              & Dec.\ ($\degr$\ $'$ $''$) at mass centre & 11:29:43.705$^{+\ <0.001}_{-\ <0.001}$     & 11:29:43.694$^{+\ <0.001}_{-0.002}$\\
    $e_{\rm x}$      & Elliptical component $x$                & $-$0.136$^{+0.002}_{-0.002}$               & $-$0.126$^{+\ <0.001}_{-0.001}$\\
    $e_{\rm y}$      & Elliptical component $y$                & $-$0.263$^{+0.002}_{-0.002}$               & $-$0.270$^{+0.003}_{-\ <0.001}$\\
    $r$              & Einstein radius ($''$)                  & 0.645$^{+0.001}_{-0.001}$                  & 0.636$^{+\ <0.001}_{-0.001}$\\
    $\alpha$         & Slope                                   & 1.897$^{+0.003}_{-0.003}$                  & 1.884$^{+\ <0.001}_{-0.014}$\\
    $\gamma_{\rm x}$ & Elliptical components $x$ of shear      & 0.034$^{+0.001}_{-0.001}$                  & 0.043$^{+\ <0.001}_{-0.004}$\\
    $\gamma_{\rm y}$ & Elliptical components $y$ of shear      & $-$0.054$^{+0.001}_{-0.001}$               & $-$0.057$^{+\ <0.001}_{-0.003}$\\
    \hline
    \end{tabular}
    \caption{Best fitted parameters with $1-\sigma$ confidence of the lens object models derived from the ALMA 300-GHz continuum data.}
    \label{tab:parameters of lens}
\end{table*}

\subsection{Non-parametric fitting}

Non-parametric source reconstructions with \texttt{PyAutoLens} are performed
using a Voronoi mesh, where the Voronoi cells adapt to the source morphology.
In this way, a relatively higher resolution is dedicated to its bright central
regions than that in the outer region with weaker signals.  The source
reconstruction is performed after an assumed parametric mass model maps
coordinates from the lens plane to the source plane with ray-tracing.

\subsubsection{The ALMA 300-GHz continuum }

When modelling the 300-GHz sub-millimetre data, we adopt iterative steps.  To
fit the non-parametric source, we first fix the parameters of the lens object
according to our parametric mass model (see Sect.~\ref{para_models}), and then
use a pixelized Voronoi grid that adapts to the mass model's magnification
pattern, giving higher resolutions to areas with higher magnification.

Then, we perform a fit where the source plane is reconstructed by a
brightness-weighted Voronoi (Fig.~\ref{fig:SourceImages} left) mesh. We fit a
new mass model, a power-law ellipsoid \citep{Tessore2015} with an external
shear (shown in Table \ref{tab:parameters of lens}). With the Bayesian
inference (Dynesty) integrated within PyAutoLens, we adopt the best-fit results
and their associated errors as the lens profile (Table \ref{tab:parameters of
lens}), which is further used for the non-parametric modelling of the radio
data. 

\subsubsection{The 1.5-, 8.4-, and 33-GHz continuum}

The ALMA 300-GHz continuum image provides strong constraints on our lensing
galaxy mass model, therefore we fix the mass model parameters to values derived
from this data in order to perform non-parametric fitting at radio bands. Using
the same mass model across all wavelengths also ensures source reconstructions
can be compared in a uniform way.

For the data from VLA and e-Merlin, the source plane is reconstructed with a
magnification-weighted Voronoi mesh instead of a brightness-weighted one.
This is because a magnification-based mesh produces an identical grid of
Voronoi cells across all three frequency channels, ensuring consistent sampling
of the source when calculating the spectral indices pixel by pixel to obtain a
spectral indices map.  The residual maps at multiple wavelengths are presented
in Appendix \ref{App:Residual}.  We measure the radio fluxes within two
regions in the source plane, shown as white and red ellipses in Figure
\ref{fig:SourceImages}. The error compromises a 15\% absolute flux uncertainty,
a ${\sim}\,5\%$ error from the uncertainty of the model by running multiple
models with all parameters randomly sampled with their probability density
distributions, and a statistical error measured in emission-free regions.  The
measured quantities of different components in the reconstructed source plane
are listed in Table \ref{tab:source profile}.

To verify the consistency between the optical data and our
parametric/non-parametric models, we invert a simple point  source to the lens
plane, with lens mass models derived from parametric  and non-parametric
fitting.  Both positions and fluxes of the point source are adopted from the
result first fitted with the HST data. For both parametric and non-parametric
models, the positions of the simulated  point source (Fig.~\ref{fig:dataimage})
match well with all four images in the HST data, within 0.5-pixel size.  This
indicates that our models are reliable. 

\begin{table*}
    \centering
    \begin{tabular}{lcccc}
    \hline
    Component    & Frequency & Flux                                   & Intrinsic Size                       & Peak Position \\
                 & (GHz)     & (\textmu Jy)                           & ($\arcsec \times \arcsec$) & (R.A., Dec.)\\
    \hline
    host galaxy  & 300       & 2.0 $\pm$ 0.3 [$\pm$ 0.3] $\times10^3$ & $0.12 \times 0.11$         & (14:15:46.2378, 11:29:43.659)\\
    host galaxy  & 33        & 4.8 $\pm$ 1.3 [$\pm$ 0.7]              & $0.14 \times 0.11$         & (14:15:46.2364, 11:29:43.659)\\
    host galaxy  & 8.4         & 7 $\pm$ 2 [$\pm$ 1]              & $0.14 \times 0.095$         & (14:15:46.2383, 11:29:43.698)\\
    host galaxy  & 1.5       & 26 $\pm$ 10 [$\pm$ 4]                  & -                          & (14:15:46.2388, 11:29:43.689)\\
    \hline
    radio jet   & 33        & 4.7 $\pm$ 1.0 [$\pm$ 0.7]              & $0.15 \times 0.086$        & (14:15:46.2315, 11:29:43.787)\\
    radio jet   & 8.4         & 14 $\pm$ 3 [$\pm$ 2]              & $0.11 \times 0.049$        & (14:15:46.2315, 11:29:43.781)\\
    radio jet   & 1.5       & 1.1 $\pm$ 0.2 [$\pm$ 0.2] $\times10^2$ & $0.12 \times 0.045$        & (14:15:46.2299, 11:29:43.781)\\
    \hline
    \end{tabular}
    \caption{Measured quantities of the Cloverleaf components in the source
            plane at different wavelengths, where the fluxes are obtained
            within the ellipses in Fig.~\ref{fig:SourceImages}, and the
            intrinsic sizes are described in (the FWHM of the major axis)
            $\times$ (the FWHM of the minor axis). The error (1-$\sigma$) consists of
    contributions from statistical error and model uncertainty and from the
    absolute calibration error (15\,\%, shown in the squared bracket). } 
    \label{tab:source profile}
\end{table*}

\section{Results} \label{sec:result}

\subsection{Spatial distributions of multi-wavelength images } 

As shown in Sect.~\ref{sec:model}, the HST optical images are point sources
dominated by the central AGN.  On the other hand, the 300 GHz image (see
Fig.~\ref{fig:dataimage} top right) shows that all four components are much larger
than the beam size (see Table \ref{tab:images table}), indicating that the
sub-mm continuum emission has extended structures. 

%

The radio continuum images (shown in Fig.~\ref{fig:dataimage}) also show resolved
extended structures.  The B and D components tend to merge when moving to the
lower frequency, indicating that the source is lying on the caustics. The other
two, components A and C, both also have spatial deviations, not only between
different radio bands but also between the images from HST and ALMA.

At 1.5 GHz band, the radio-to-submm spatial offsets are $(0.24\arcsec,
-0.01\arcsec)$ and $(-0.03\arcsec, 0.24\arcsec)$ in R.A.\ and Dec., for
components A and C, respectively.  All components in the images of 8.4\,GHz and
33\,GHz are irregular with respect to the synthesized beam shapes at the
corresponding wavelengths; thus, we do not fit their locations with {\sc imfit}.

Among the three radio bands (1.5\,GHz, 8.4\,GHz and 33\,GHz), the image
morphologies are not identical in the source plane. The 1.5\,GHz image shows almost no emission
from the host galaxy, while both 8.4 and 33\,GHz radio images show detections at
the same position.  This suggests the multiple
energy sources among different wavelengths.

\subsection{The discovery of a radio jet }

As shown in the reconstructed image of the source plane at 300\,GHz  of
Cloverleaf (Fig.~\ref{fig:SourceImages}), the cold dust emission shows an
extended disk-like structure of $0.115''\times0.109''$  in FWHM
($0.946\,\mathrm{kpc}\times0.897\,\mathrm{kpc}$) rather than a point source.

\begin{figure*}
    \centering
    \includegraphics[scale=0.3]{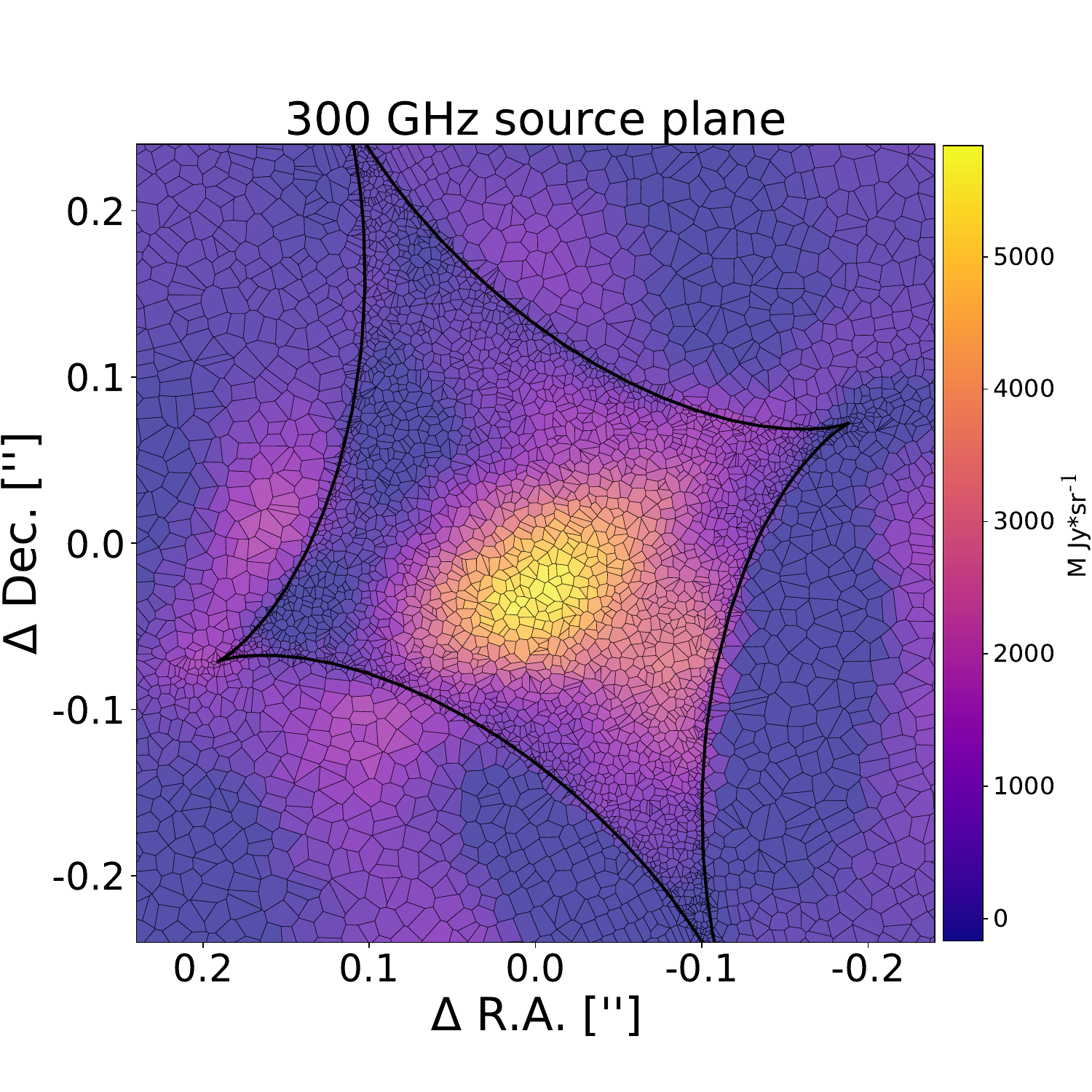}
    \includegraphics[scale=0.3]{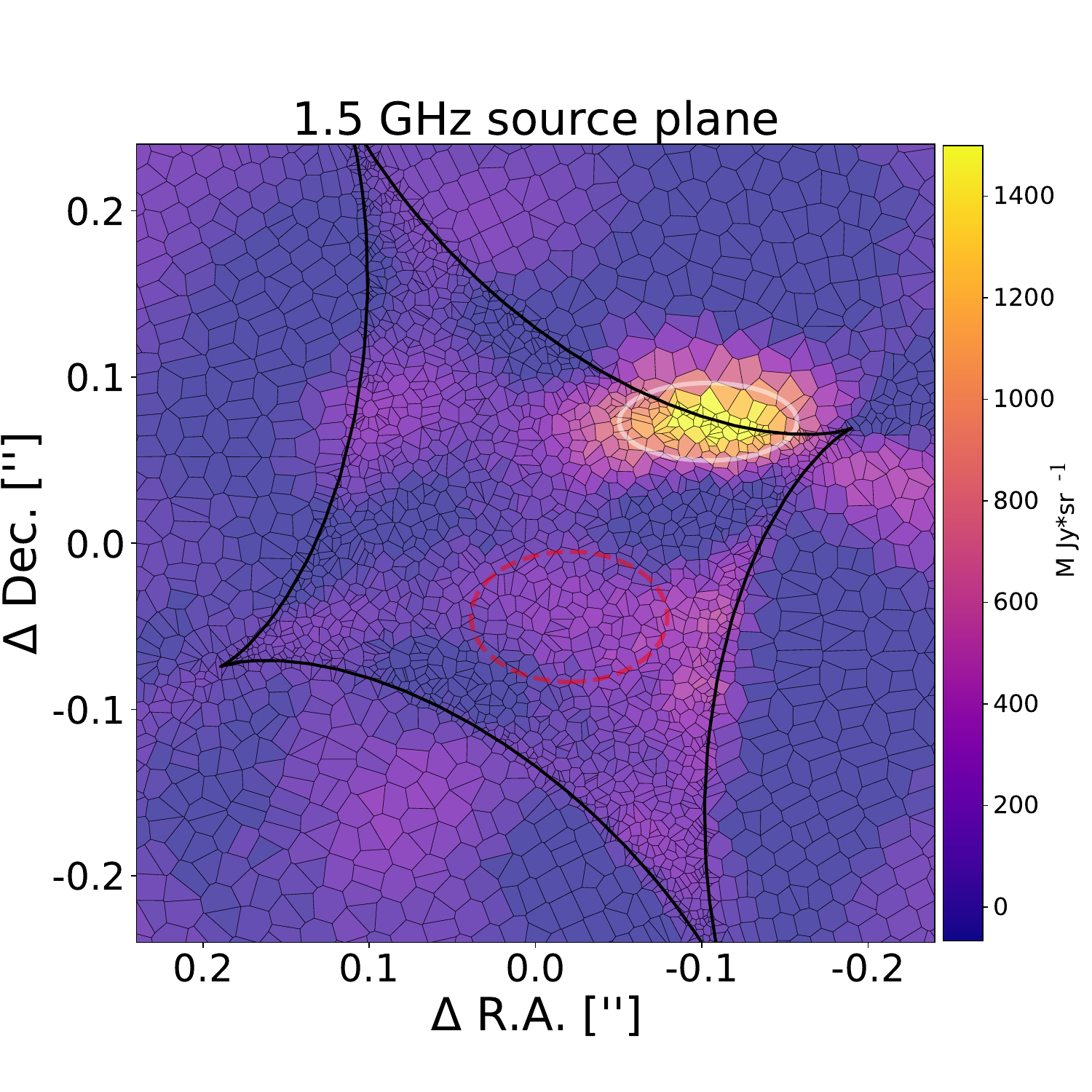}
    \includegraphics[scale=0.3]{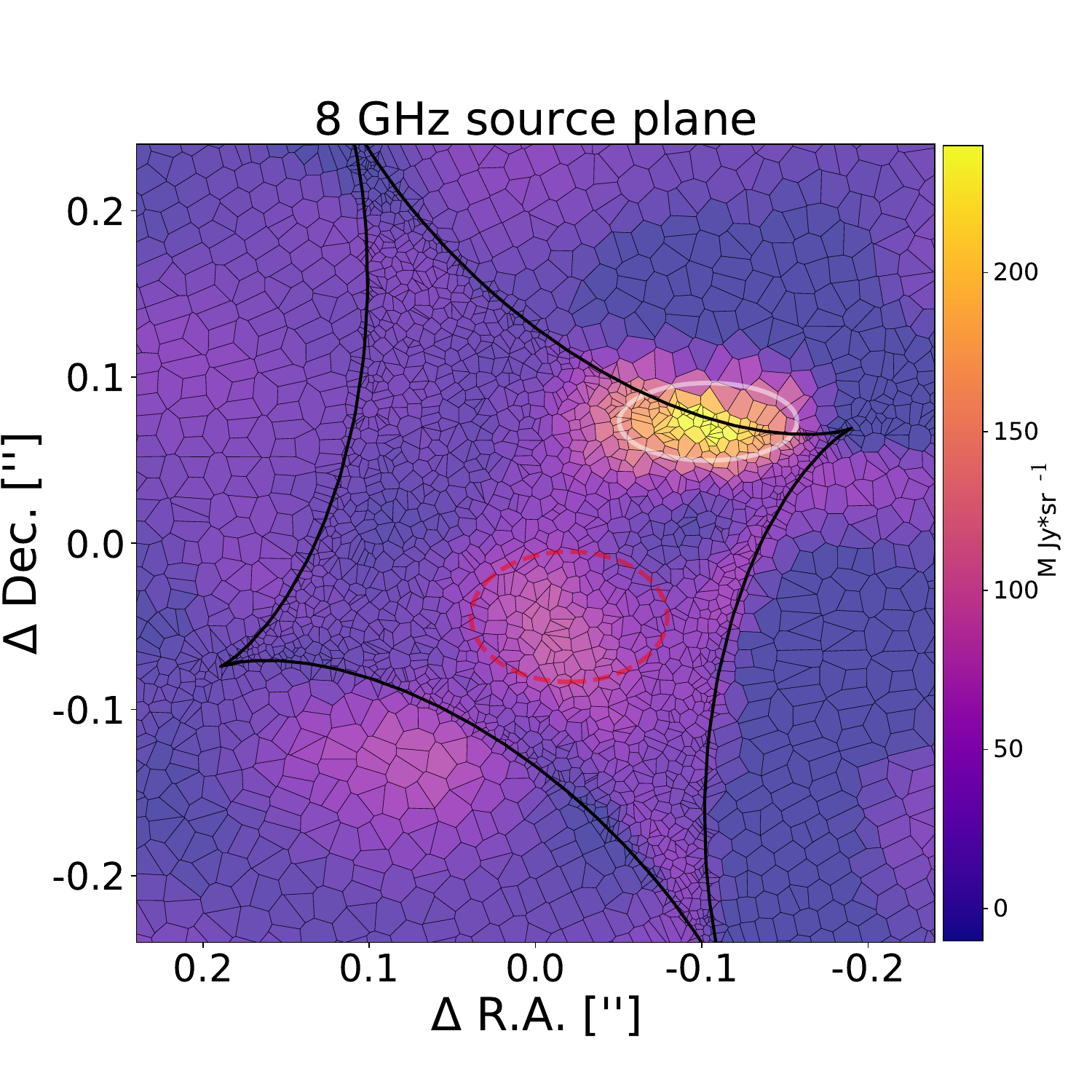}
    \includegraphics[scale=0.3]{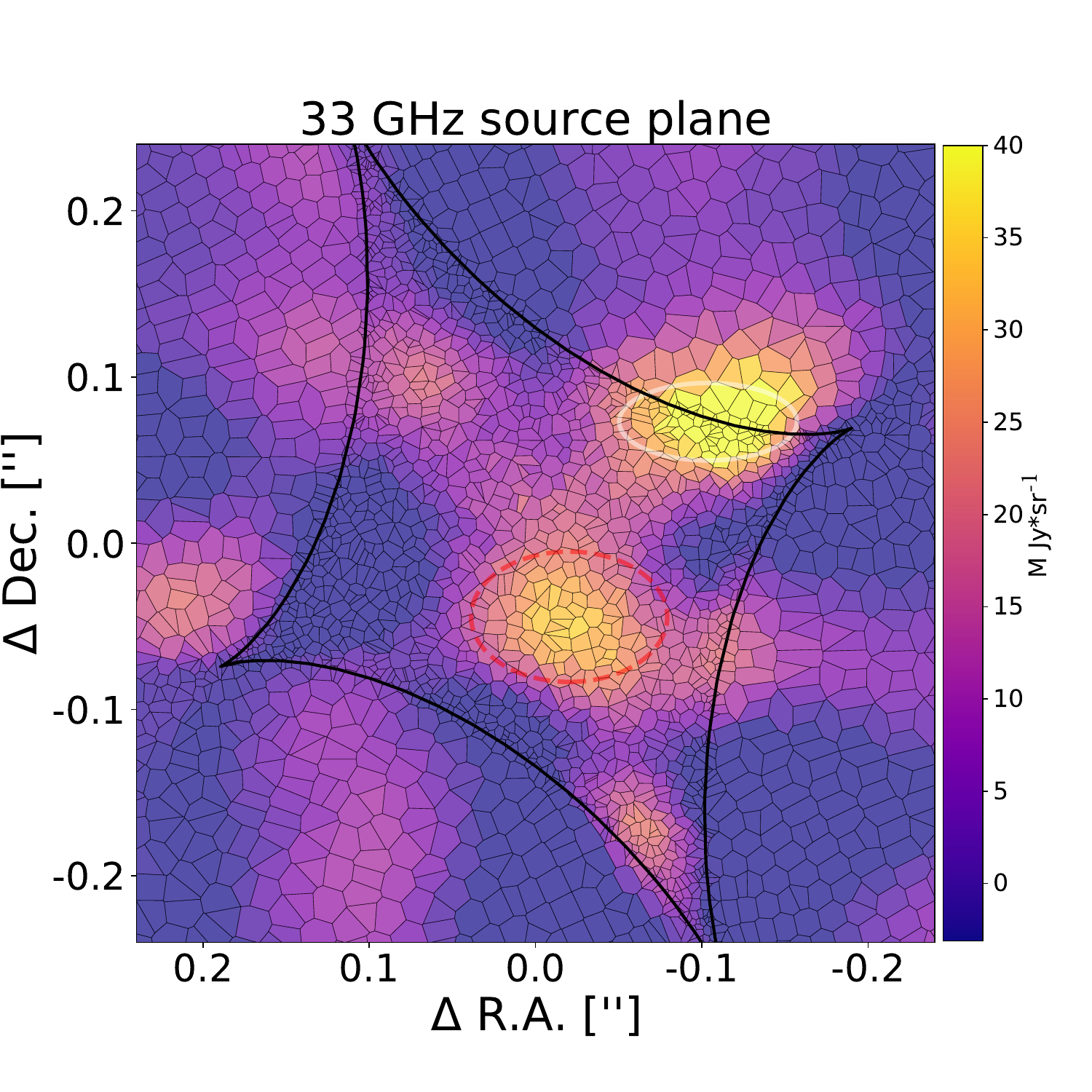}
    \caption{Reconstructed source planes of Cloverleaf at multiple bands: top
    left: sub-millimetre continuum at 300\,GHz; top right: radio continuum at 1.5\,GHz;
    bottom left: radio continuum at 8.4\,GHz; bottom right: radio continuum at
    33\,GHz. The solid black line is the inner caustics. The centre is on
    (14:15:46.$2384^\mathrm{h}$, 11:29:43.705) in the J2000.0 equatorial system of coordinates.
    The white solid ellipse and the red dashed ellipse are the areas to draw
    fluxes for SEDs of the radio jet and the host galaxy, respectively. }
    \label{fig:SourceImages}
\end{figure*}

The peak of this dust continuum has an S/N $\sim 27$, with sub-structures
around it. The majority of the host galaxy emission is distributed in the
central region, inside the inner caustics, while a small clump is located in
the southwest to the centre. Due to the uncertainty of the lens model, several
emission clumps outside the caustics have less fidelity. 

All reconstructed radio images show extended emission to the northwest of the
dust continuum core, at a projected distance of ${\sim}1.15$\,kpc
($0.14\arcsec$) away from the centre of the host galaxy. The peaks of this
radio component are roughly consistent among different radio wavebands.  This
indicates a single-sided radio jet launched from the host galaxy, with a
comparable size to the host galaxy at 300\,GHz. The FWHMs of this radio clump,
fitted with Gaussian profiles, are
$0.99\,\mathrm{kpc}\times0.37\,\mathrm{kpc}$,
$0.90\,\mathrm{kpc}\times0.40\,\mathrm{kpc}$, and
$1.23\,\mathrm{kpc}\times0.71\,\mathrm{kpc}$, at 1.5\,GHz, 8.4\,GHz, and 33\,GHz,
respectively.

On the other hand, in the source plane, the 1.5\,GHz radio emission from the host galaxy is
almost negligible, while the 8.4 GHz and 33 GHz images show radio emission at the
location of the host galaxy, with their S/Ns of 3.5 and 3.7, respectively.


\begin{figure*}
    \centering
    \includegraphics[width=0.495\textwidth]{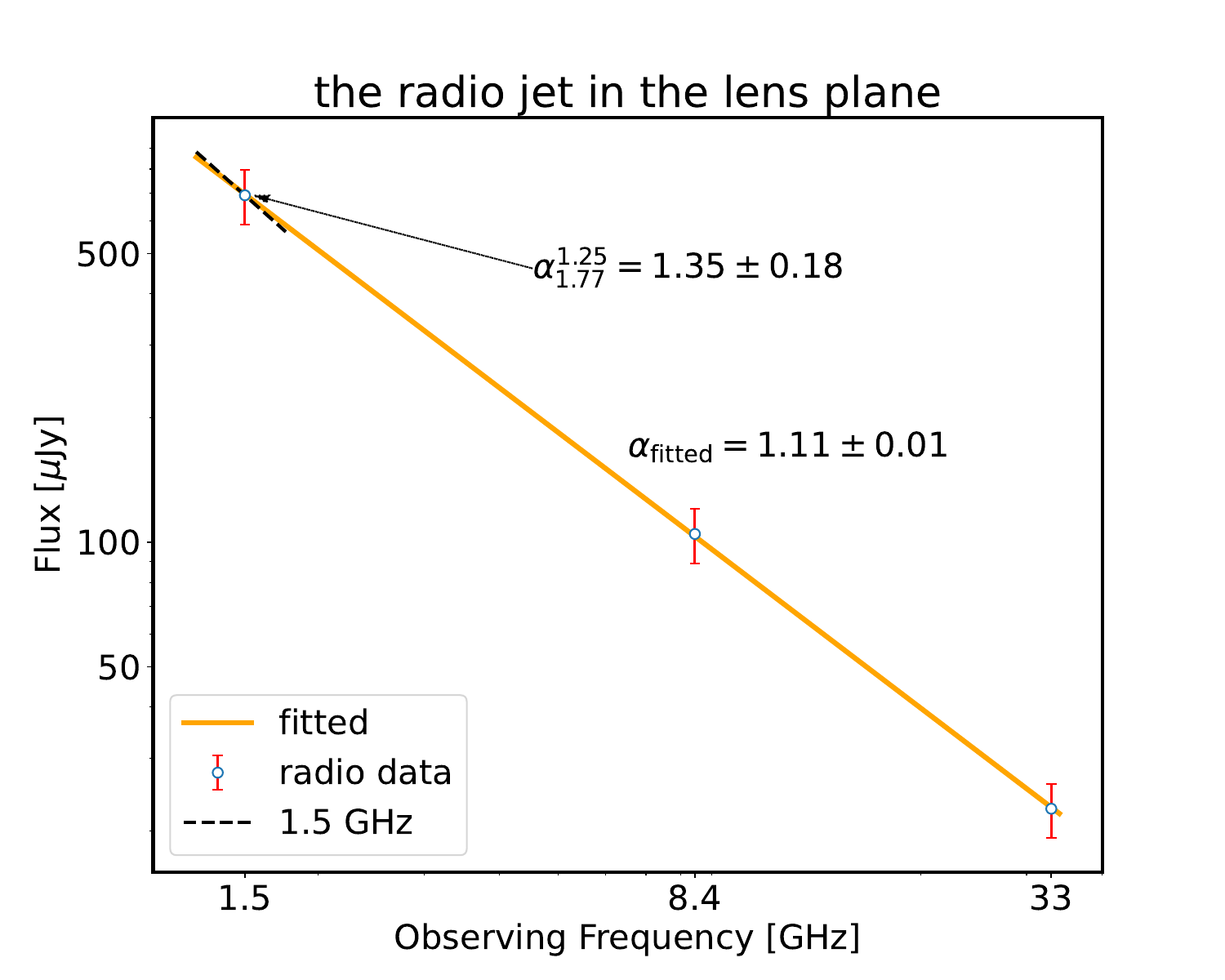}
    \includegraphics[width=0.495\textwidth]{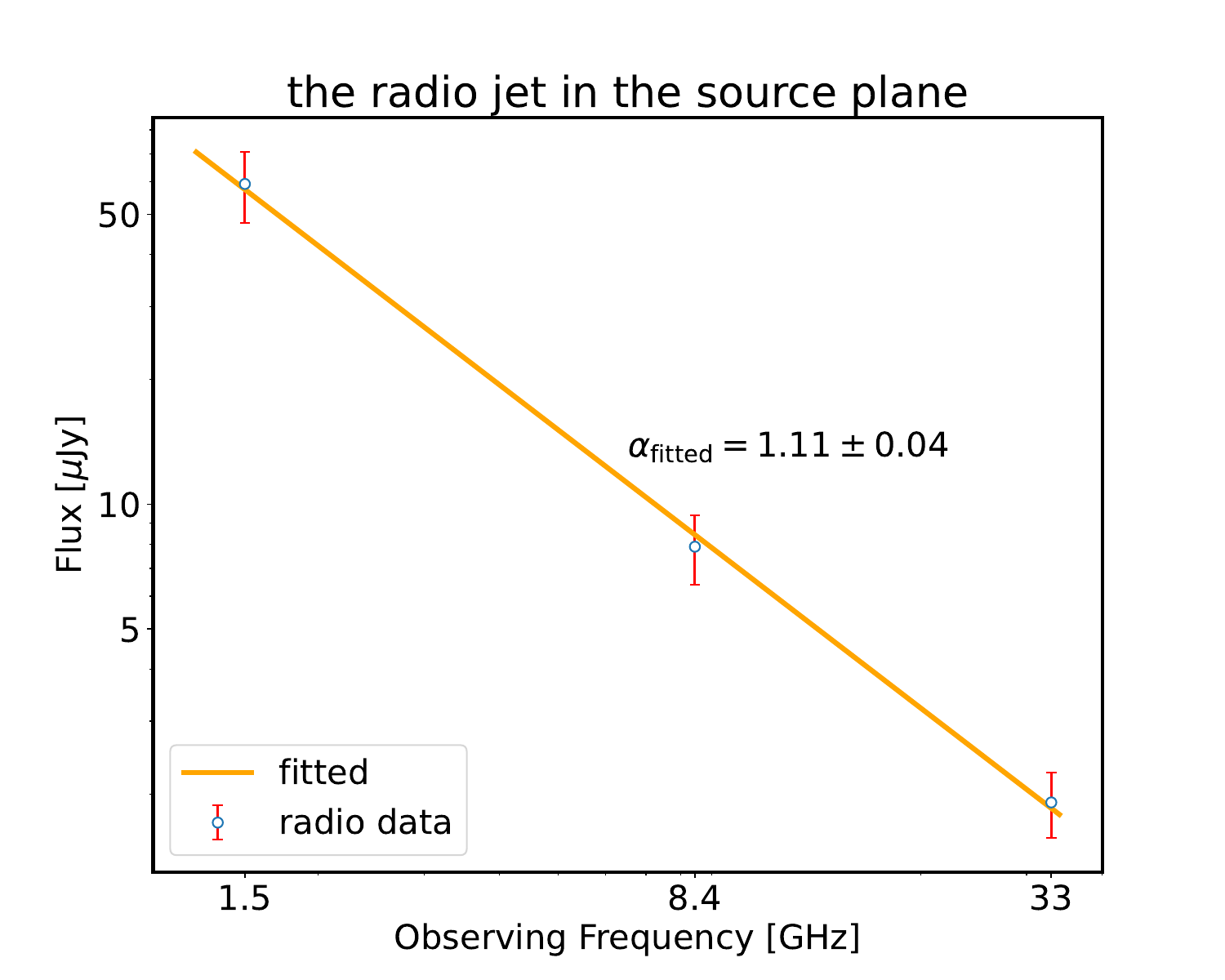}
    \includegraphics[width=0.495\textwidth]{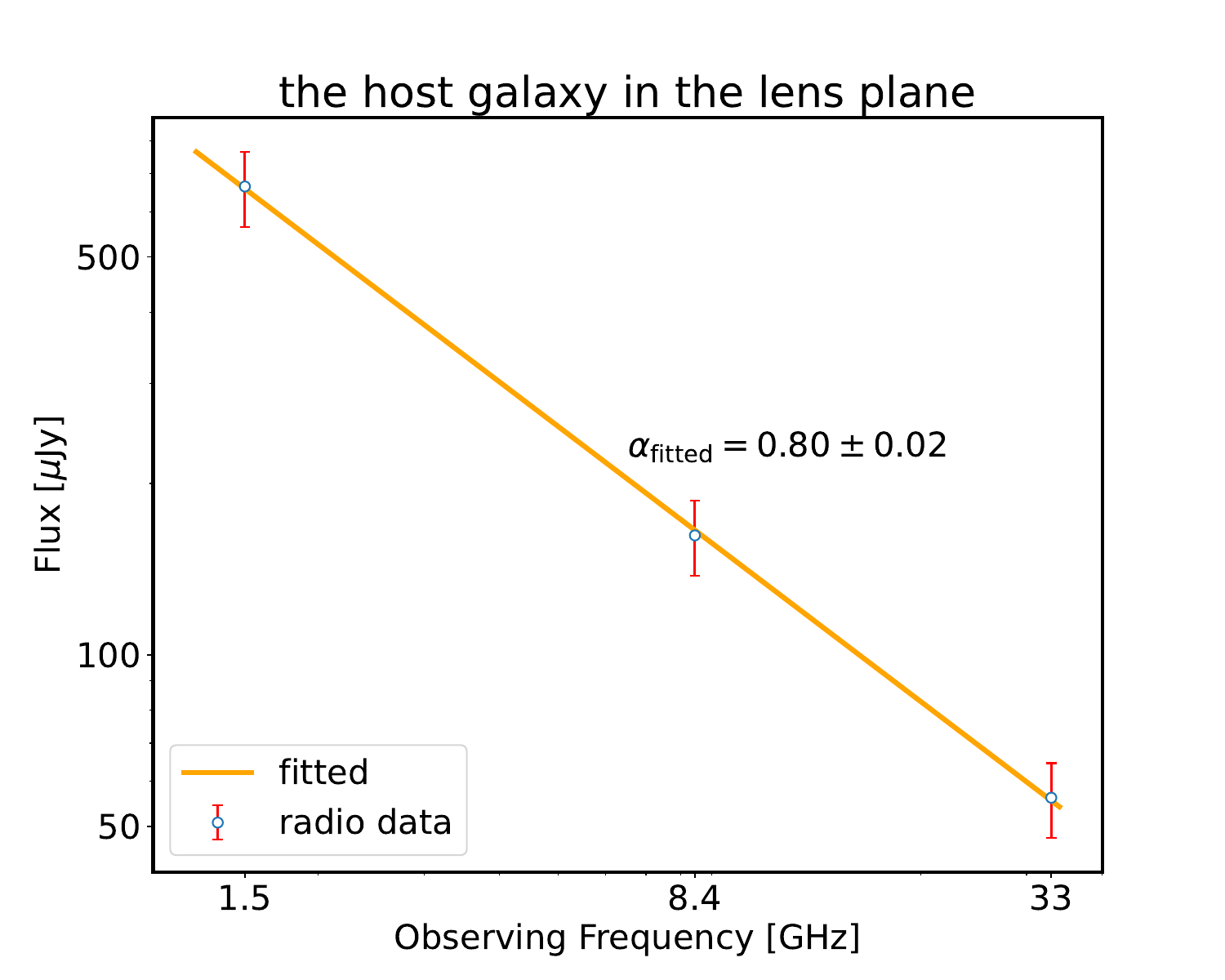}
    \includegraphics[width=0.495\textwidth]{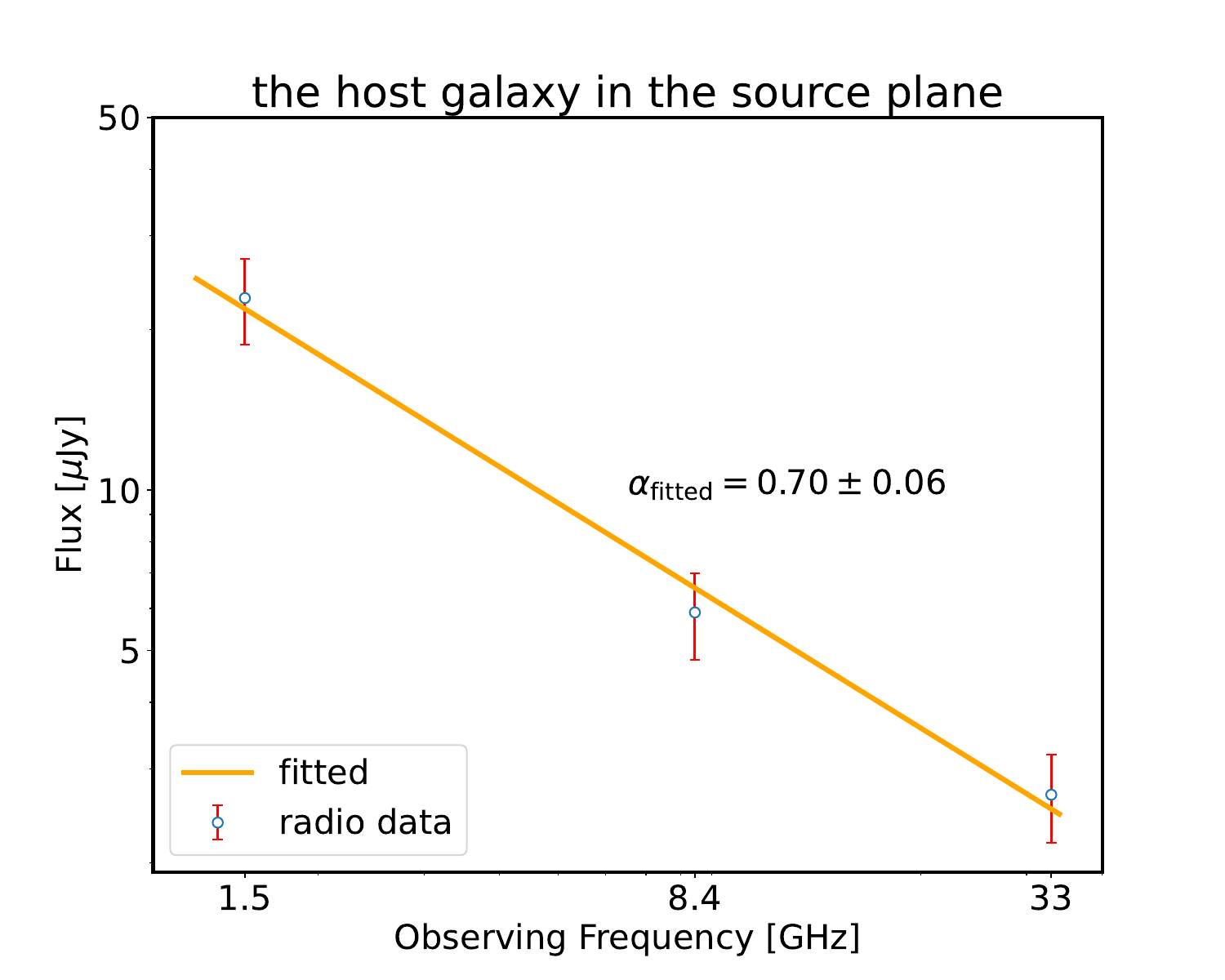}
    \caption{Spectral energy distributions (SEDs) of the radio jet and the host galaxy of 
    Cloverleaf.  {\it Top-Left}: SED of the radio jet in the lens plane. 
    {\it Top-Right}: SED of the radio jet in the source plane. 
    {\it Bottom-Left}: SED of the host galaxy in the lens plane. 
    {\it Bottom-Right}: SED of the host galaxy in the source plane. 
    Radio fluxes are measured from the regions shown in the lens plane 
    (Fig.~\ref{fig:dataimage}) and the source plane (Fig.~\ref{fig:SourceImages}). 
    Orange lines show the best linear fitting result for all three frequencies. 
    The black dashed line in the top-left panel shows the spectral index of the radio 
    jet in the lens plane derived from the 1.5 GHz e-Merlin data only. 
    All derived spectral indices are listed in Table \ref{tab:SED indices}. The error bars are at $1-\sigma$ confidence consisting of statistical error, lens model uncertainty, and absolute calibration error (15\,\%). 
    }
    \label{fig:SED}
\end{figure*}

\subsection{Spectral indices of the radio continuum}

The multiple-band radio maps allow us to probe the SEDs of both the host galaxy and the radio jet. To evaluate
the impact of the differential magnification effect and to probe the power
sources of the radio emission, we construct SEDs derived from both the lens
plane and the source plane (see Fig.~\ref{fig:SED}).  

We first measure radio fluxes from regions shown in the lens plane
(Fig.~\ref{fig:dataimage}) and the source plane (Fig.~\ref{fig:SourceImages}) for
1.5, 8.4, and 33\,GHz, respectively. 

The regions adopted to measure fluxes in the source plane are shown in Fig.~\ref{fig:SourceImages}. 
In the lens plane, however, the emission from the host galaxy and the radio jet do not show distinct spatial offsets between each other.  The radio
emission of components B and D tend to merge together, indicating that this
emission component is lying on the caustics (Appendix \ref{App:Mock}), where the radio jet populates while the host galaxy does not.  Therefore, to eliminate
contamination from the host galaxy, we choose this region (shown as black
ellipses in Fig.~\ref{fig:dataimage}) in the lens plane and construct the SED of
the lensed radio structure.  On the other hand, we adopt beam-sized regions at the
peaks of the four image components to measure fluxes from the host galaxy.  We
then sum up fluxes from all four components to increase the S/N.

Then we fit spectral indices with fluxes from all three radio bands, which are
shown in orange lines of Fig.~\ref{fig:SED}. All SEDs can be fitted with one single power-law profile.

\begin{figure*}
    \centering
    \includegraphics[scale=0.4]{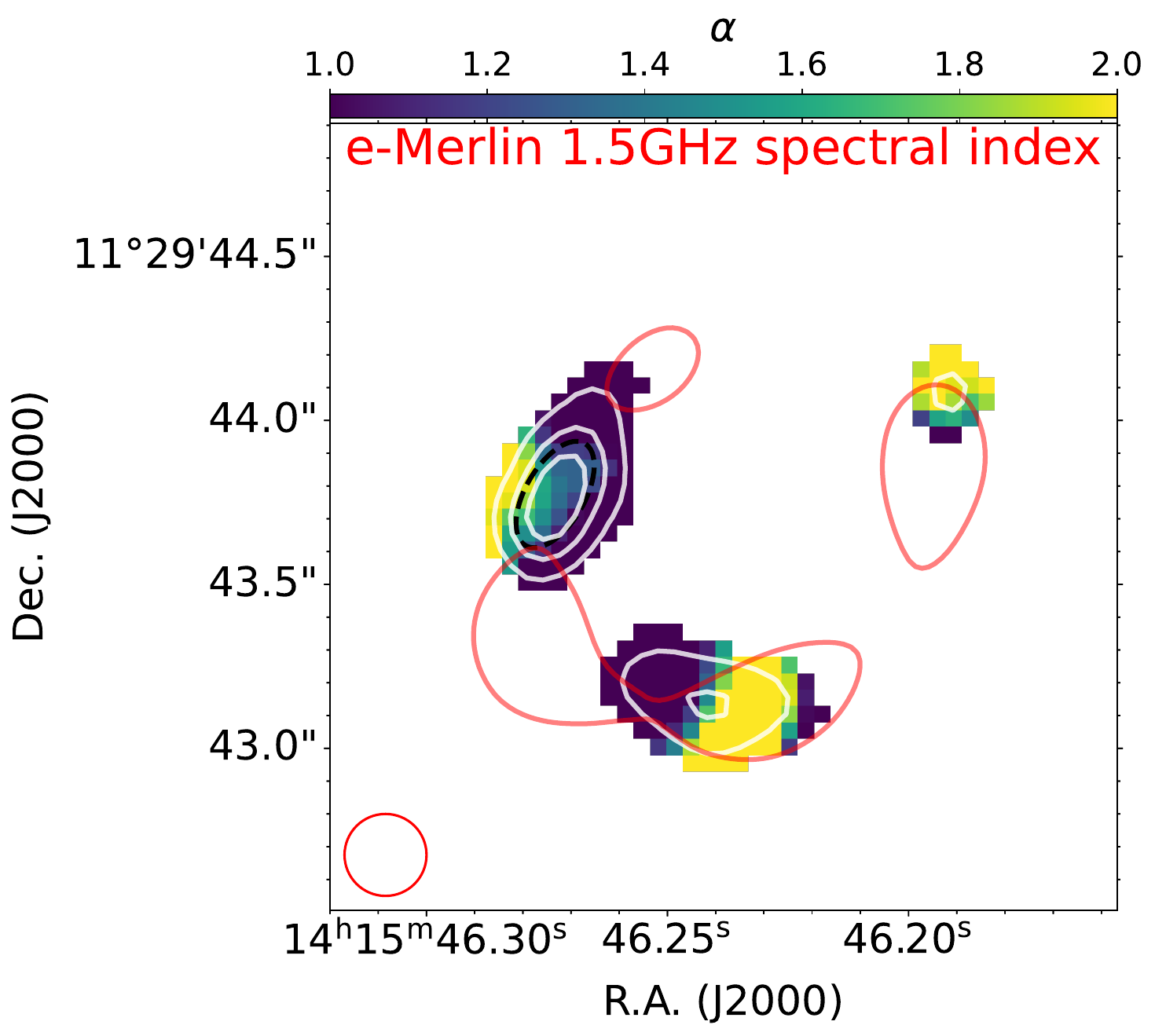}
    \includegraphics[scale=0.4]{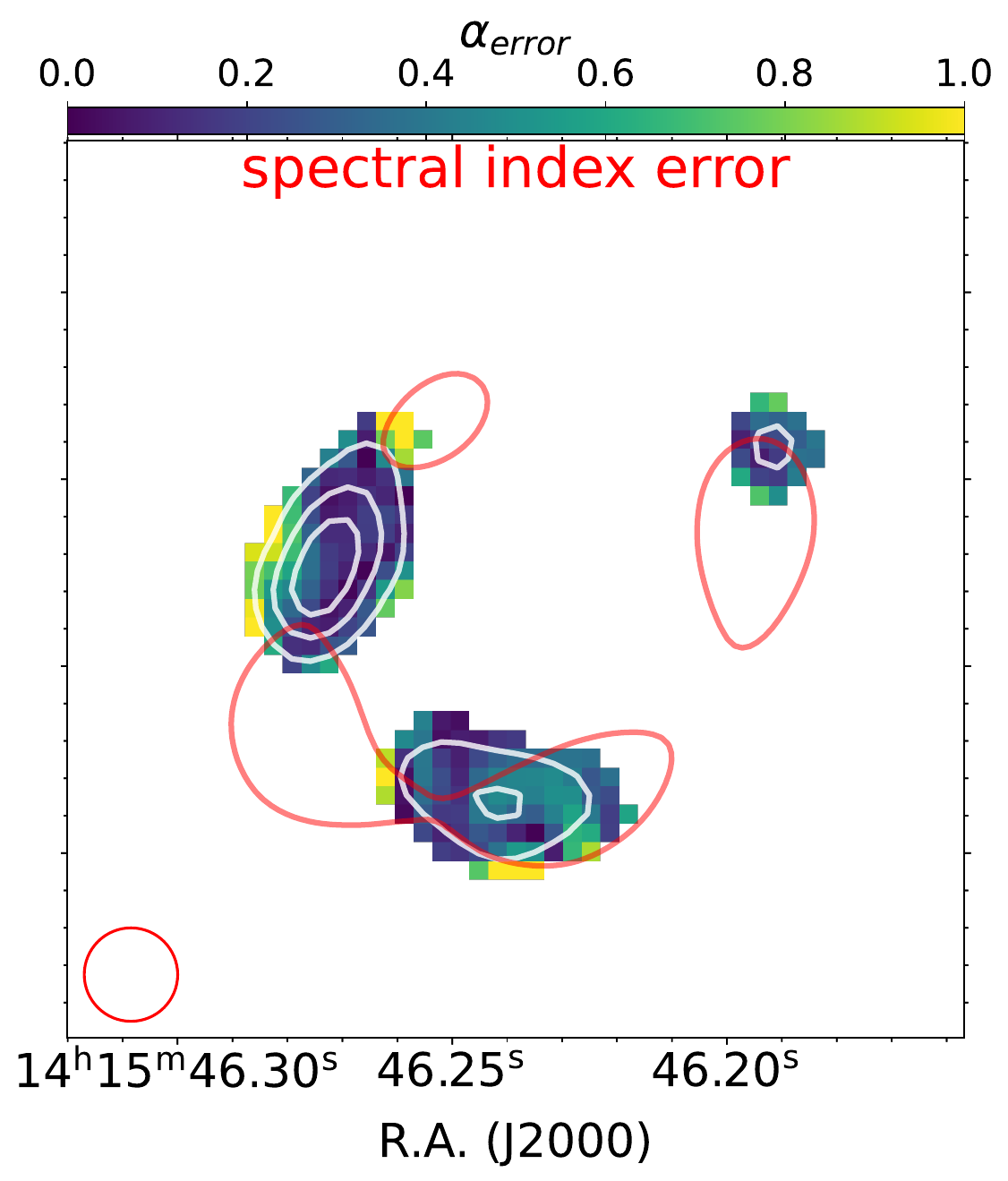}
    \caption{Spectral index map (Left) and its error map (Right) in the lens
    plane at 1.5 GHz. The white and red lines are the contours of the images
    at 1.5 GHz ($8-, 14-, 20-\sigma$) and 300 GHz ($50-\sigma$), respectively. The red circles represent the beam
    sizes. The mask we use is a 5-$\sigma$ cutoff from the e-Merlin continuum. The black dashed line is the area we derive the spectral index at 1.5 GHz. }
    \label{fig:localspectralindex}
\end{figure*}

We also use the e-Merlin data only, which covers 1.25 -- 1.77\,GHz bandwidth,
to fit a spectral index map (Fig.~\ref{fig:localspectralindex}).  
We use a 5-$\sigma$ mask from the e-Merlin continuum to mask the spectral index map and the error map. 
The average spectral index of the merging components B and D, which are labeled with the black dashed ellipse shown in Fig. \ref{fig:localspectralindex}, same as the black ellipse in Fig. \ref{fig:dataimage}, 
to be $\alpha ^{1.25}_{1.77}=1.35 \pm 0.18$. It represents the average spectral index of the lensed radio jet at 1.5\,GHz in the lens plane.
All derived spectral indices for both the host galaxy and the radio jet, are
listed in Table~\ref{tab:SED indices}.

Although the average slope of the e-Merlin spectral index map is consistent
with that obtained with $\alpha^{1.5}_{8.4}$ within 1-$\sigma$ in the whole region, it also shows a
systematic gradient of the radio slope from the east to the west direction.
This indicates that the radio jet may have complex structures or physical
processes that need higher-resolution data to study in detail.

\begin{figure}
    \centering
    \includegraphics[width=0.5\textwidth]{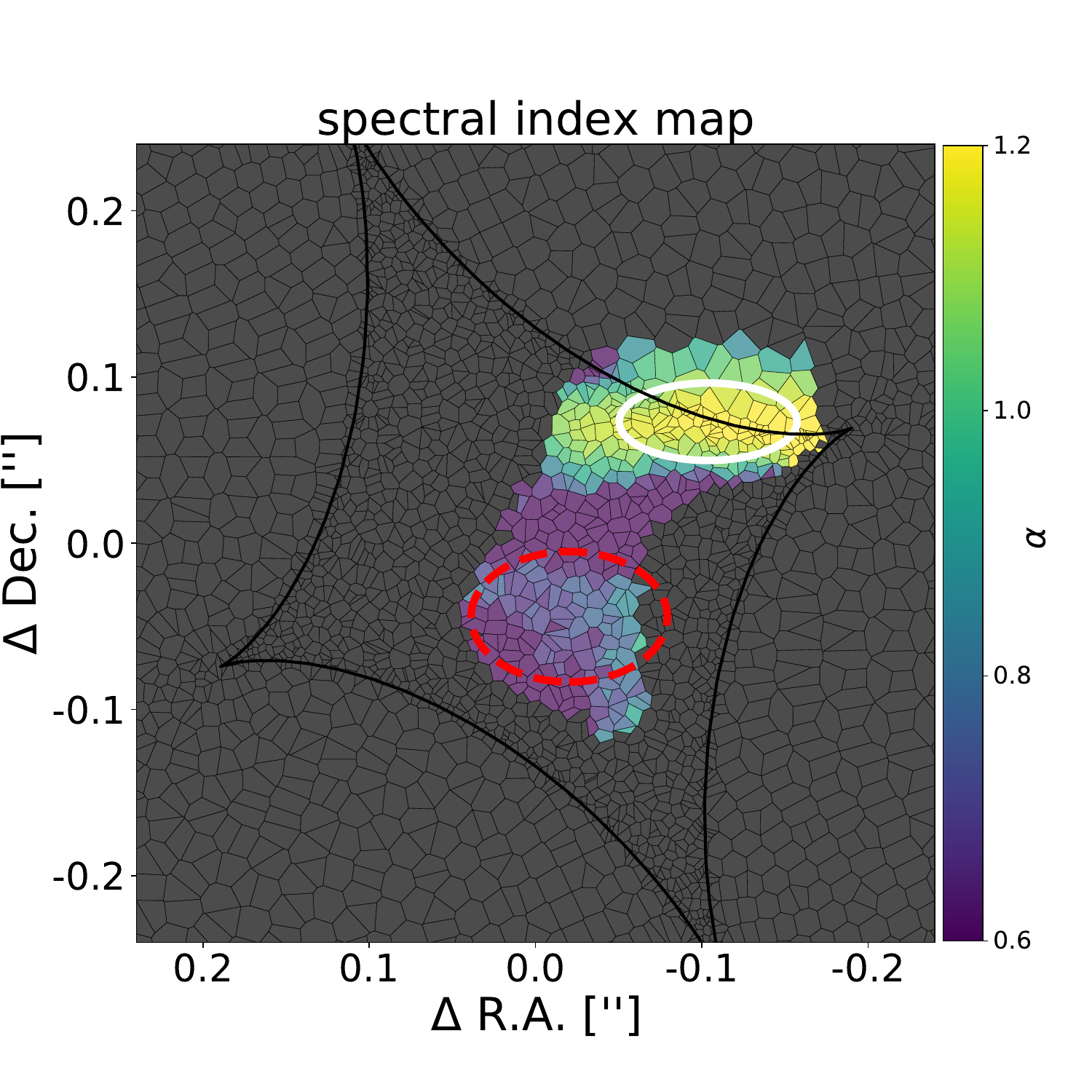}
    \caption{Spectral index map (derived from the reconstructed images at 1.5-, 8.4-, and 33-\,GHz) in the source plane: The black solid line is
    the inner caustics. The centre is on (14:15:46.$2384^h$, 11:29:43.705) in
    the J2000 equatorial system of coordinates. The white solid ellipse and the red dashed ellipse mark the positions of the radio jet and the host galaxy, respectively. } \label{fig:SpectralIndex}
\end{figure}

From the flux maps shown in Fig.~\ref{fig:SourceImages}, we fit a spectral
index map (Fig.~\ref{fig:SpectralIndex}, by the opposite sign convention of
$\alpha$) using de-lensed flux maps in the source plane.  As shown in
Fig.~\ref{fig:SpectralIndex}, we fit power-law profiles for all pixels at maps of 8.4\,GHz, 1.5\,GHz, and 33\,GHz, with a mask of $\mathrm{S/N} > 3.5$ at the 33\,GHz source plane image.  
The spectral index
map shows flatter SED at the host galaxy position, while steeper SED at the
radio jet.

\begin{table*}
    \centering
    \begin{tabular}{cccccc}
    \hline
    Component   & Plane  & $\alpha ^{1.5}_{8.4}$ & $\alpha ^{8.4}_{33}$ & $\alpha _{\rm{fitted}}$ & $\alpha ^{1.25}_{1.77}$ \\
    \hline
    radio jet  & lens   & $1.10 \pm 0.12$     & $1.12 \pm 0.16$    & $1.11 \pm 0.01$ & $1.35 \pm 0.18$ \\
    radio jet  & source & $1.17 \pm 0.16$     & $1.04 \pm 0.19$    & $1.11 \pm 0.04$ & - \\
    host galaxy & lens   & $0.84 \pm 0.12$     & $0.76 \pm 0.16$    & $0.80 \pm 0.02$ & - \\
    host galaxy & source & $0.79 \pm 0.15$     & $0.58 \pm 0.19$   & $0.70 \pm 0.06$ & - \\
    \hline
    \end{tabular}
    \caption{Spectral indices of different components in the lens plane and the
            source plane within 1-$\sigma$: $\alpha _{\rm fitted}$ is the spectral index of the
    best linear fit, considering fluxes in all three radio bands. }
    \label{tab:SED indices}
\end{table*}

\section{Discussion} \label{sec:discuss}

\subsection{Position offsets between sub-millimetre and radio emission}

Compared with the lensed image of Cloverleaf at $\sim$300\,GHz, those at radio
bands have different spatial distributions, 
which indicates that the emission
distributions at different bands are different from one another in the source
plane, similar to the case of strongly lensed quasar MG J0751+2716 with a radio jet \citep{Spingola2018, Powell2022}.  This can be inferred directly from the reconstructed source planes (see
Fig.~\ref{fig:SourceImages}).  All reconstructed radio images contain radio
structures on scales of ${\sim}0.9\times0.4\,\mathrm{kpc}$, on the northwest
part of the inner caustics, which are absent at the sub-millimetre wavelength.

\subsection{Possibility to be a jet in the foreground?}

One may also ask if the radio feature comes from the foreground lens galaxy,
instead of the background source galaxy.  First, the multiple components shown
in all radio bands indicate that it is unlikely to be multiple jets from the
centre. Second, the spatial distributions of all radio components are highly
consistent with the scenario that the radio structure locates on the caustics,
i.e. merging between lensed components (Appendix \ref{App:Mock}).  Therefore,
we suggest that this radio structure should reside in the source plane and
originate from the position on the northwest part of caustics.

\subsection{Why One-sided?}

On the other hand, the radio source only presents on one side of the host galaxy, 
while the other side shows almost no signal. Depending on the geometry, the
other side of the radio jet might be too faint to be detected due to being away from the caustics, 
or simply does not exist. In any case, if the other side of the radio jet exists,  it should
lead to emission populating the area between components A and C (see Appendix
\ref{App:Mock}). Future deeper observations would be the key to verifying this
scenario.

\subsection{The Power-Law Profiles of the SEDs}

As shown in Fig.~\ref{fig:SED} and Table~\ref{tab:SED indices}, the SEDs of the
radio jet in both lens plane and source plane exhibit power-law profiles, and have similar spectral indices, which indicates that the emission is dominated by the synchrotron from electrons in the jet material. 

On the position of the host galaxy, the SED can be also fitted with one single power-law profile with a flatter spectral index (see Fig. \ref{fig:SpectralIndex} and Table \ref{tab:SED indices}) compared with the one at the position of the radio jet. 
The emission (rest frequency, $f_{\rm rest} \sim$5-120\,GHz) at this position may have a contribution from free-free and/or dust continuum \citep{Condon1992}, together with a contribution from the synchrotron emitted from the relativistic electron accelerated by supernovae remnants. 

\subsection{AGN Feedback of Cloverleaf} 

Based on the broad absorption line profile \citep{Hazard1984, Magain1988} and the HST optical images (Fig. \ref{fig:dataimage}) of this quasar, the wind/radiation feedback could take place in this galaxy.

On the other hand, this radio jet/structure is located at a projected distance of ${\sim}1.15$\,kpc
($0.14\arcsec$) from the galaxy centre, with a size similar to that of the host
galaxy at 33\,GHz. 
These are consistent with
radio lobe features found in massive radio galaxies \citep{Fabian2012}, 
however, Cloverleaf seems to have a lower galaxy mass than those massive
galaxies \citep{Solomon2003}.
The kinetic energy residing in the jet could severely heat up the gas and drive strong feedback to the star formation in Cloverleaf.

\citet{Chartas2004} constrained the mass of the SMBH in the Cloverleaf quasar by the microlensed image A (the one located at the south in the four images shown in Fig. \ref{fig:dataimage}) at X-ray band. They also adopted the luminosity from \citet{Granato1996}'s dust-enshrouded AGN models. Then the Eddington ratio of this AGN is found to be $\sim 0.1$, assuming the magnification of the AGN to be 11 \citep{Venturini2003}. The Eddington ratio of 0.1 indicates that the AGN is not radiating close to the Eddington limit, yet is still a powerful quasar.

Therefore, given its strong ongoing starburst, likely driven by the interaction with a neighbouring
gas-rich galaxy \citep{Stacey2022}, together with the broad absorption line and the newly discovered radio jet, we suggest that the AGN feedback in Cloverleaf consists of coexistent contributions from both radiation-driven wind and AGN-launched jet. 
High angular resolution observations
of both radio continuum and molecular gas tracers in the future, which help
resolve the jet structure and gas kinematics, could throw light on the
co-existence of both mechanisms of AGN feedback.



\section{Conclusion} \label{sec:conclusion}

With high angular resolution photometric data at optical, sub-mm, and radio
wavelengths, we reconstructed multi-band source images of the gravitational
lensed Cloverleaf QSO, with both parametric and non-parametric modelling.  From
different emission locations, we discovered a single-sided radio jet, with
a size of $\sim$1\,kpc, located at projected $\sim$1.2\,kpc to the northwest of Cloverleaf. 
The co-existence of a radio jet and broad absorption line indicates that the Cloverleaf quasar likely feedbacks its host galaxy with both AGN wind and radio jet. 
This provides a unique case
for the galaxy-SMBH co-evolution studies at high redshifts. 

\section*{Acknowledgements}

Based on observations made with the NASA/ESA Hubble Space Telescope, and
obtained from the Hubble Legacy Archive, which is a collaboration between the
Space Telescope Science Institute (STScI/NASA), the Space Telescope European
Coordinating Facility (ST-ECF/ESAC/ESA) and the Canadian Astronomy Data Centre
(CADC/NRC/CSA). 
LZ acknowledges the referee for pointing out the mistake in data reduction and for kind and helpful suggestions, which heavily polish this study. 
LZ appreciates Yunwei Deng and Chengjiang Yin for their technical helps. 
ZYZ and LZ acknowledge support of the National Natural Science Foundation of China (NSFC) under grants No. 12041305, 12173016.  
ZYZ and LZ acknowledge the Program for Innovative Talents, Entrepreneur in Jiangsu.  
ZYZ and LZ acknowledge the science research grants from the China Manned Space Project with NO.CMS-CSST-2021-A08. 
CY acknowledges support from ERC Advanced Grant 789410. 
DDX acknowledges the NSFC Grant 12073013. 
This work is supported by the China Manned Space Project (No. CMS-CSST-2021-A07).
RL acknowledges the support of National Nature Science Foundation of China (Nos 11988101,11773032,12022306), 
the support from the Ministry of Science and Technology of China (Nos. 2020SKA0110100),  
the science research grants from the China Manned Space Project (Nos.\ CMS-CSST-2021-B01,CMS-CSST-2021-A01), 
CAS Project for Young Scientists in Basic Research(No.\ YSBR-062).
Z.-C.\ Z.\ is supported by the National Natural Science Foundation of China (Grant No.\ 12233002).

This work used the following software: Autolens\citep{Nightingale2015, Nightingale2018, pyautolens} , CASA
\citep{McMullin2007}, SAOimageDS9 \citep{Joye2003}, astropy
\citep{astropy1,astropy2}, dynesty \citep{dynesty}, corner \citep{corner},
emcee \citep{emcee}, matplotlib \citep{matplotlib}, numpy \citep{numpy},
pyautofit\citep{pyautofit}, pylops\citep{pylops},  python \citep{python}, scipy
\citep{scipy}

\section*{Data Availability}

All data adopted in this work are publicly available in corresponding data archival systems of HST, ALMA, VLA, and e-Merlin.



\bibliographystyle{mnras}
\bibliography{references} 

@ARTICLE{Fabian2012,
       author = {{Fabian}, A.~C.},
        title = "{Observational Evidence of Active Galactic Nuclei Feedback}",
      journal = {\araa},
         year = 2012,
        month = sep,
       volume = {50},
        pages = {455-489},
          doi = {10.1146/annurev-astro-081811-125521},
archivePrefix = {arXiv},
       eprint = {1204.4114},
 primaryClass = {astro-ph.CO},
       adsurl = {https://ui.adsabs.harvard.edu/abs/2012ARA&A..50..455F}
}

@ARTICLE{Kormendy2013,
       author = {{Kormendy}, John and {Ho}, Luis C.},
        title = "{Coevolution (Or Not) of Supermassive Black Holes and Host Galaxies}",
      journal = {\araa},
         year = 2013,
        month = aug,
       volume = {51},
       number = {1},
        pages = {511-653},
          doi = {10.1146/annurev-astro-082708-101811},
archivePrefix = {arXiv},
       eprint = {1304.7762},
 primaryClass = {astro-ph.CO},
       adsurl = {https://ui.adsabs.harvard.edu/abs/2013ARA&A..51..511K}
}

@ARTICLE{Churazov2005,
       author = {{Churazov}, E. and {Sazonov}, S. and {Sunyaev}, R. and {Forman}, W. and {Jones}, C. and {B{\"o}hringer}, H.},
        title = "{Supermassive black holes in elliptical galaxies: switching from very bright to very dim}",
      journal = {\mnras},
         year = 2005,
        month = oct,
       volume = {363},
       number = {1},
        pages = {L91-L95},
          doi = {10.1111/j.1745-3933.2005.00093.x},
archivePrefix = {arXiv},
       eprint = {astro-ph/0507073},
 primaryClass = {astro-ph},
       adsurl = {https://ui.adsabs.harvard.edu/abs/2005MNRAS.363L..91C}
}

@ARTICLE{Solomon2003,
       author = {{Solomon}, P. and {Vanden Bout}, P. and {Carilli}, C. and {Guelin}, M.},
        title = "{The essential signature of a massive starburst in a distant quasar}",
      journal = {\nat},
         year = 2003,
        month = dec,
       volume = {426},
       number = {6967},
        pages = {636-638},
          doi = {10.1038/nature02149},
archivePrefix = {arXiv},
       eprint = {astro-ph/0312436},
 primaryClass = {astro-ph},
       adsurl = {https://ui.adsabs.harvard.edu/abs/2003Natur.426..636S}
}

@ARTICLE{Venturini2003,
       author = {{Venturini}, S. and {Solomon}, P.~M.},
        title = "{The Molecular Disk in the Cloverleaf Quasar}",
      journal = {\apj},
         year = 2003,
        month = jun,
       volume = {590},
       number = {2},
        pages = {740-745},
          doi = {10.1086/375050},
archivePrefix = {arXiv},
       eprint = {astro-ph/0210529},
 primaryClass = {astro-ph},
       adsurl = {https://ui.adsabs.harvard.edu/abs/2003ApJ...590..740V}
}

@ARTICLE{Barvainis1997,
       author = {{Barvainis}, Richard and {Maloney}, Philip and {Antonucci}, Robert and {Alloin}, Danielle},
        title = "{Multiple CO Transitions, C I, and HCN from the Cloverleaf Quasar}",
      journal = {\apj},
         year = 1997,
        month = jul,
       volume = {484},
       number = {2},
        pages = {695-701},
          doi = {10.1086/304382},
archivePrefix = {arXiv},
       eprint = {astro-ph/9702118},
 primaryClass = {astro-ph},
       adsurl = {https://ui.adsabs.harvard.edu/abs/1997ApJ...484..695B}
}

@ARTICLE{Weiss2003,
       author = {{Wei{\ss}}, A. and {Henkel}, C. and {Downes}, D. and {Walter}, F.},
        title = "{Gas and dust in the Cloverleaf quasar at redshift 2.5}",
      journal = {\aap},
         year = 2003,
        month = oct,
       volume = {409},
        pages = {L41-L45},
          doi = {10.1051/0004-6361:20031337},
archivePrefix = {arXiv},
       eprint = {astro-ph/0309048},
 primaryClass = {astro-ph},
       adsurl = {https://ui.adsabs.harvard.edu/abs/2003A&A...409L..41W}
}

@ARTICLE{Kneib1998,
       author = {{Kneib}, J. -P. and {Alloin}, D. and {Mellier}, Y. and {Guilloteau}, S. and {Barvainis}, R. and {Antonucci}, R.},
        title = "{Modelling the Cloverleaf: contribution of a galaxy cluster at Z \raisebox{-0.5ex}\textasciitilde 1.7}",
      journal = {\aap},
         year = 1998,
        month = jan,
       volume = {329},
        pages = {827-839},
archivePrefix = {arXiv},
       eprint = {astro-ph/9706036},
 primaryClass = {astro-ph},
       adsurl = {https://ui.adsabs.harvard.edu/abs/1998A&A...329..827K}
}

@INPROCEEDINGS{McMullin2007,
       author = {{McMullin}, J.~P. and {Waters}, B. and {Schiebel}, D. and {Young}, W. and {Golap}, K.},
        title = "{CASA Architecture and Applications}",
    booktitle = {Astronomical Data Analysis Software and Systems XVI},
         year = 2007,
       editor = {{Shaw}, R.~A. and {Hill}, F. and {Bell}, D.~J.},
       series = {Astronomical Society of the Pacific Conference Series},
       volume = {376},
        month = oct,
        pages = {127},
       adsurl = {https://ui.adsabs.harvard.edu/abs/2007ASPC..376..127M}
}

@article{pyautolens,
  doi = {10.21105/joss.02825},
  url = {https://doi.org/10.21105/joss.02825},
  year = {2021},
  publisher = {The Open Journal},
  volume = {6},
  number = {58},
  pages = {2825},
  author = {Nightingale, J. W. and Hayes, R. G. and Ashley Kelly and Aristeidis Amvrosiadis and Amy Etherington and Qiuhan He and Nan Li and XiaoYue Cao and Jonathan Frawley and Shaun Cole and Andrea Enia and Carlos S. Frenk and David R. Harvey and Ran Li and Richard J. Massey and Mattia Negrello and Andrew Robertson},
  title = {`PyAutoLens`: Open-Source Strong Gravitational Lensing},
  journal = {J. Open Source Softw.}
}

@article{Nightingale2015,
        archivePrefix = {arXiv},
        arxivId = {1412.7436},
        author = {Nightingale, J. W. and Dye, S.},
        doi = {10.1093/mnras/stv1455},
        eprint = {1412.7436},
        issn = {13652966},
        journal = {MNRAS},
        month = {sep},
        number = {3},
        pages = {2940--2959},
        title = {{Adaptive semi-linear inversion of strong gravitational lens imaging}},
        volume = {452},
        year = {2015}
}

@article{Nightingale2018,
        archivePrefix = {arXiv},
        arxivId = {1708.07377},
        author = {Nightingale, J. W. and Dye, S. and Massey, Richard J.},
        doi = {10.1093/mnras/sty1264},
        eprint = {1708.07377},
        issn = {13652966},
        journal = {MNRAS},
        number = {4},
        pages = {4738--4784},
        title = {{AutoLens: Automated modeling of a strong lens's light, mass, and source}},
        url = {https://academic.oup.com/mnras/article/478/4/4738/5001434},
        volume = {478},
        year = {2018}
}

@INPROCEEDINGS{Bertin2002,
       author = {{Bertin}, Emmanuel and {Mellier}, Yannick and {Radovich}, Mario and {Missonnier}, Gilles and {Didelon}, Pierre and {Morin}, Bertrand},
        title = "{The TERAPIX Pipeline}",
    booktitle = {Astronomical Data Analysis Software and Systems XI},
         year = 2002,
       editor = {{Bohlender}, David A. and {Durand}, Daniel and {Handley}, Thomas H.},
       series = {Astronomical Society of the Pacific Conference Series},
       volume = {281},
        month = jan,
        pages = {228},
       adsurl = {https://ui.adsabs.harvard.edu/abs/2002ASPC..281..228B}
}

@article{astropy1,
      Adsurl = {http://adsabs.harvard.edu/abs/2013A%26A...558A..33A},
      Archiveprefix = {arXiv},
      Author = {{Astropy Collaboration} and {Robitaille}, T.~P. and {Tollerud}, E.~J. and {Greenfield}, P. and {Droettboom}, M. and {Bray}, E. and {Aldcroft}, T. and {Davis}, M. and {Ginsburg}, A. and {Price-Whelan}, A.~M. and {Kerzendorf}, W.~E. and {Conley}, A. and {Crighton}, N. and {Barbary}, K. and {Muna}, D. and {Ferguson}, H. and {Grollier}, F. and {Parikh}, M.~M. and {Nair}, P.~H. and {Unther}, H.~M. and {Deil}, C. and {Woillez}, J. and {Conseil}, S. and {Kramer}, R. and {Turner}, J.~E.~H. and {Singer}, L. and {Fox}, R. and {Weaver}, B.~A. and {Zabalza}, V. and {Edwards}, Z.~I. and {Azalee Bostroem}, K. and {Burke}, D.~J. and {Casey}, A.~R. and {Crawford}, S.~M. and {Dencheva}, N. and {Ely}, J. and {Jenness}, T. and {Labrie}, K. and {Lim}, P.~L. and {Pierfederici}, F. and {Pontzen}, A. and {Ptak}, A. and {Refsdal}, B. and {Servillat}, M. and {Streicher}, O.},
      Doi = {10.1051/0004-6361/201322068},
      Eid = {A33},
      Eprint = {1307.6212},
      Journal = {A\&A},
      Month = oct,
      Pages = {A33},
      Primaryclass = {astro-ph.IM},
      Title = {{Astropy: A community Python package for astronomy}},
      Volume = 558,
      Year = 2013
}

@article{astropy2,
      Adsurl = {https://ui.adsabs.harvard.edu/#abs/2018AJ....156..123T},
      Author = {{Price-Whelan}, A.~M. and {Sip{\H{o}}cz}, B.~M. and {G{\"u}nther}, H.~M. and {Lim}, P.~L. and {Crawford}, S.~M. and {Conseil}, S. and {Shupe}, D.~L. and {Craig}, M.~W. and {Dencheva}, N. and {Ginsburg}, A. and {VanderPlas}, J.~T. and {Bradley}, L.~D. and {P{\'e}rez-Su{\'a}rez}, D. and {de Val-Borro}, M. and {Paper Contributors}, (Primary and {Aldcroft}, T.~L. and {Cruz}, K.~L. and {Robitaille}, T.~P. and {Tollerud}, E.~J. and {Coordination Committee}, (Astropy and {Ardelean}, C. and {Babej}, T. and {Bach}, Y.~P. and {Bachetti}, M. and {Bakanov}, A.~V. and {Bamford}, S.~P. and {Barentsen}, G. and {Barmby}, P. and {Baumbach}, A. and {Berry}, K.~L. and {Biscani}, F. and {Boquien}, M. and {Bostroem}, K.~A. and {Bouma}, L.~G. and {Brammer}, G.~B. and {Bray}, E.~M. and {Breytenbach}, H. and {Buddelmeijer}, H. and {Burke}, D.~J. and {Calderone}, G. and {Cano Rodr{\'\i}guez}, J.~L. and {Cara}, M. and {Cardoso}, J.~V.~M. and {Cheedella}, S. and {Copin}, Y. and {Corrales}, L. and {Crichton}, D. and {D{\textquoteright}Avella}, D. and {Deil}, C. and {Depagne}, {\'E}. and {Dietrich}, J.~P. and {Donath}, A. and {Droettboom}, M. and {Earl}, N. and {Erben}, T. and {Fabbro}, S. and {Ferreira}, L.~A. and {Finethy}, T. and {Fox}, R.~T. and {Garrison}, L.~H. and {Gibbons}, S.~L.~J. and {Goldstein}, D.~A. and {Gommers}, R. and {Greco}, J.~P. and {Greenfield}, P. and {Groener}, A.~M. and {Grollier}, F. and {Hagen}, A. and {Hirst}, P. and {Homeier}, D. and {Horton}, A.~J. and {Hosseinzadeh}, G. and {Hu}, L. and {Hunkeler}, J.~S. and {Ivezi{\'c}}, {\v{Z}}. and {Jain}, A. and {Jenness}, T. and {Kanarek}, G. and {Kendrew}, S. and {Kern}, N.~S. and {Kerzendorf}, W.~E. and {Khvalko}, A. and {King}, J. and {Kirkby}, D. and {Kulkarni}, A.~M. and {Kumar}, A. and {Lee}, A. and {Lenz}, D. and {Littlefair}, S.~P. and {Ma}, Z. and {Macleod}, D.~M. and {Mastropietro}, M. and {McCully}, C. and {Montagnac}, S. and {Morris}, B.~M. and {Mueller}, M. and {Mumford}, S.~J. and {Muna}, D. and {Murphy}, N.~A. and {Nelson}, S. and {Nguyen}, G.~H. and {Ninan}, J.~P. and {N{\"o}the}, M. and {Ogaz}, S. and {Oh}, S. and {Parejko}, J.~K. and {Parley}, N. and {Pascual}, S. and {Patil}, R. and {Patil}, A.~A. and {Plunkett}, A.~L. and {Prochaska}, J.~X. and {Rastogi}, T. and {Reddy Janga}, V. and {Sabater}, J. and {Sakurikar}, P. and {Seifert}, M. and {Sherbert}, L.~E. and {Sherwood-Taylor}, H. and {Shih}, A.~Y. and {Sick}, J. and {Silbiger}, M.~T. and {Singanamalla}, S. and {Singer}, L.~P. and {Sladen}, P.~H. and {Sooley}, K.~A. and {Sornarajah}, S. and {Streicher}, O. and {Teuben}, P. and {Thomas}, S.~W. and {Tremblay}, G.~R. and {Turner}, J.~E.~H. and {Terr{\'o}n}, V. and {van Kerkwijk}, M.~H. and {de la Vega}, A. and {Watkins}, L.~L. and {Weaver}, B.~A. and {Whitmore}, J.~B. and {Woillez}, J. and {Zabalza}, V. and {Contributors}, (Astropy)},
      Doi = {10.3847/1538-3881/aabc4f},
      Eid = {123},
      Journal = {AJ},
      Month = Sep,
      Pages = {123},
      Primaryclass = {astro-ph.IM},
      Title = {{The Astropy Project: Building an Open-science Project and Status of the v2.0 Core Package}},
      Volume = {156},
      Year = 2018
}

@article{corner,
      doi = {10.21105/joss.00024},
      url = {https://doi.org/10.21105/joss.00024},
      year  = {2016},
      month = {jun},
      publisher = {The Open Journal},
      volume = {1},
      number = {2},
      pages = {24},
      author = {Daniel Foreman-Mackey},
      title = {corner.py: Scatterplot matrices in Python},
      journal = {The J. Open Source Softw.}
}

@article{dynesty,
      archivePrefix = {arXiv},
      arxivId = {1904.02180},
      author = {Speagle, Joshua S},
      doi = {10.1093/mnras/staa278},
      eprint = {1904.02180},
      issn = {0035-8711},
      journal = {MNRAS},
      number = {3},
      pages = {3132--3158},
      title = {{dynesty: a dynamic nested sampling package for estimating Bayesian posteriors and evidences}},
      volume = {493},
      year = {2020}
}

@article{emcee,
     archivePrefix = {arXiv},
     arxivId = {1202.3665},
     author = {Foreman-Mackey, Daniel and Hogg, David W. and Lang, Dustin and Goodman, Jonathan},
     doi = {10.1086/670067},
     eprint = {1202.3665},
     issn = {00046280},
     journal = {Publ. Astron. Soc. Pac.},
     number = {925},
     pages = {306--312},
     title = {{emcee : The MCMC Hammer }},
     volume = {125},
     year = {2013}
}

@article{matplotlib,
  Author    = {Hunter, J. D.},
  Title     = {Matplotlib: A 2D graphics environment},
  Journal   = {Comput Sci Eng},
  Volume    = {9},
  Number    = {3},
  Pages     = {90--95},
  publisher = {IEEE COMPUTER SOC},
  doi       = {10.1109/MCSE.2007.55},
  year      = 2007
}

@article{numpy,
  author={S. {van der Walt} and S. C. {Colbert} and G. {Varoquaux}},
  doi={10.1109/MCSE.2011.37},
  journal={Comput Sci Eng},
  title={The NumPy Array2D: A Structure for Efficient Numerical Computation},
  year={2011},
  volume={13},
  number={2},
  pages={22-30},}

@article{pyautofit,
  doi = {10.21105/joss.02550},
  url = {https://doi.org/10.21105/joss.02550},
  year = {2021},
  publisher = {The Open Journal},
  volume = {6},
  number = {58},
  pages = {2550},
  author = {Nightingale, J. W. and Hayes, R. G. and Griffiths, M.},
  title = {`PyAutoFit`: A Classy Probabilistic Programming Language for Model Composition and Fitting},
  journal = {J. Open Source Softw.}
}

@article{PyLops,
    title = {PyLops—A linear-operator Python library for scalable algebra and optimization},
    journal = {SoftwareX},
    volume = {11},
    pages = {100361},
    year = {2020},
    issn = {2352-7110},
    doi = {https://doi.org/10.1016/j.softx.2019.100361},
    url = {https://www.sciencedirect.com/science/article/pii/S2352711019301086},
    author = {Matteo Ravasi and Ivan Vasconcelos},
    archivePrefix = {arXiv},
    arxivId = {1907.12349},
    eprint = {1907.12349}
}

@book{python,
 author = {Van Rossum, Guido and Drake, Fred L.},
 title = {Python 3 Reference Manual},
 year = {2009},
 isbn = {1441412697},
 publisher = {CreateSpace},
 address = {Scotts Valley, CA}
}

@article{scipy,
      author = {{Virtanen}, Pauli and {Gommers}, Ralf and {Oliphant}, Travis E. and {Haberland}, Matt and {Reddy}, Tyler and {Cournapeau}, David and {Burovski}, Evgeni and {Peterson}, Pearu and {Weckesser}, Warren and {Bright}, Jonathan and {van der Walt}, St{\'e}fan J.  and {Brett}, Matthew and {Wilson}, Joshua and {Jarrod Millman}, K.  and {Mayorov}, Nikolay and {Nelson}, Andrew R.~J. and {Jones}, Eric and {Kern}, Robert and {Larson}, Eric and {Carey}, CJ and {Polat}, {\.I}lhan and {Feng}, Yu and {Moore}, Eric W. and {Vand erPlas}, Jake and {Laxalde}, Denis and {Perktold}, Josef and {Cimrman}, Robert and {Henriksen}, Ian and {Quintero}, E.~A. and {Harris}, Charles R and {Archibald}, Anne M.  and {Ribeiro}, Ant{\^o}nio H. and {Pedregosa}, Fabian and {van Mulbregt}, Paul and {Contributors}, SciPy 1. 0}, 
      title = "{SciPy 1.0: Fundamental Algorithms for Scientific Computing in Python}", journal = {Nature Methods}, 
      year = "2020",
      volume={17},
      pages={261--272},
      adsurl = {https://rdcu.be/b08Wh},
      doi = {10.1038/s41592-019-0686-2},
}

@ARTICLE{Ferrarese2005,
       author = {{Ferrarese}, Laura and {Ford}, Holland},
        title = "{Supermassive Black Holes in Galactic Nuclei: Past, Present and Future Research}",
      journal = {\ssr},
         year = 2005,
        month = feb,
       volume = {116},
       number = {3-4},
        pages = {523-624},
          doi = {10.1007/s11214-005-3947-6},
archivePrefix = {arXiv},
       eprint = {astro-ph/0411247},
 primaryClass = {astro-ph},
       adsurl = {https://ui.adsabs.harvard.edu/abs/2005SSRv..116..523F}
}

@ARTICLE{Schmidt1983,
       author = {{Schmidt}, M. and {Green}, R.~F.},
        title = "{Quasar evolution derived from the Palomar bright quasar survey and other complete quasar surveys.}",
      journal = {\apj},
         year = 1983,
        month = jun,
       volume = {269},
        pages = {352-374},
          doi = {10.1086/161048},
       adsurl = {https://ui.adsabs.harvard.edu/abs/1983ApJ...269..352S}
}

@ARTICLE{Wolf2003,
       author = {{Wolf}, C. and {Wisotzki}, L. and {Borch}, A. and {Dye}, S. and {Kleinheinrich}, M. and {Meisenheimer}, K.},
        title = "{The evolution of faint AGN between z  =\raisebox{-0.5ex}\textasciitilde  1 and z =\raisebox{-0.5ex}\textasciitilde 5 from the COMBO-17 survey}",
      journal = {\aap},
         year = 2003,
        month = sep,
       volume = {408},
        pages = {499-514},
          doi = {10.1051/0004-6361:20030990},
archivePrefix = {arXiv},
       eprint = {astro-ph/0304072},
 primaryClass = {astro-ph},
       adsurl = {https://ui.adsabs.harvard.edu/abs/2003A&A...408..499W}
}

@ARTICLE{Dunn2010,
       author = {{Dunn}, Jay P. and {Bautista}, Manuel and {Arav}, Nahum and {Moe}, Max and {Korista}, Kirk and {Costantini}, Elisa and {Benn}, Chris and {Ellison}, Sara and {Edmonds}, Doug},
        title = "{The Quasar Outflow Contribution to AGN Feedback: VLT Measurements of SDSS J0318-0600}",
      journal = {\apj},
         year = 2010,
        month = feb,
       volume = {709},
       number = {2},
        pages = {611-631},
          doi = {10.1088/0004-637X/709/2/611},
archivePrefix = {arXiv},
       eprint = {0911.3896},
 primaryClass = {astro-ph.CO},
       adsurl = {https://ui.adsabs.harvard.edu/abs/2010ApJ...709..611D}
}

@ARTICLE{Moe2009,
       author = {{Moe}, Maxwell and {Arav}, Nahum and {Bautista}, Manuel A. and {Korista}, Kirk T.},
        title = "{Quasar Outflow Contribution to AGN Feedback: Observations of QSO SDSS J0838+2955}",
      journal = {\apj},
         year = 2009,
        month = nov,
       volume = {706},
       number = {1},
        pages = {525-534},
          doi = {10.1088/0004-637X/706/1/525},
archivePrefix = {arXiv},
       eprint = {0911.3332},
 primaryClass = {astro-ph.CO},
       adsurl = {https://ui.adsabs.harvard.edu/abs/2009ApJ...706..525M}
}

@ARTICLE{Saez2009,
       author = {{Saez}, C. and {Chartas}, G. and {Brandt}, W.~N.},
        title = "{Suzaku Observations of Near-Relativistic Outflows in the BAL Quasar APM 08279+5255}",
      journal = {\apj},
         year = 2009,
        month = may,
       volume = {697},
       number = {1},
        pages = {194-206},
          doi = {10.1088/0004-637X/697/1/194},
archivePrefix = {arXiv},
       eprint = {0903.2878},
 primaryClass = {astro-ph.CO},
       adsurl = {https://ui.adsabs.harvard.edu/abs/2009ApJ...697..194S}
}

@ARTICLE{Farrah2012,
       author = {{Farrah}, Duncan and {Urrutia}, Tanya and {Lacy}, Mark and {Efstathiou}, Andreas and {Afonso}, Jose and {Coppin}, Kristen and {Hall}, Patrick B. and {Lonsdale}, Carol and {Jarrett}, Tom and {Bridge}, Carrie and {Borys}, Colin and {Petty}, Sara},
        title = "{Direct Evidence for Termination of Obscured Star Formation by Radiatively Driven Outflows in Reddened QSOs}",
      journal = {\apj},
         year = 2012,
        month = feb,
       volume = {745},
       number = {2},
          eid = {178},
        pages = {178},
          doi = {10.1088/0004-637X/745/2/178},
archivePrefix = {arXiv},
       eprint = {1112.1092},
 primaryClass = {astro-ph.CO},
       adsurl = {https://ui.adsabs.harvard.edu/abs/2012ApJ...745..178F}
}

@ARTICLE{Cano2012,
       author = {{Cano-D{\'\i}az}, M. and {Maiolino}, R. and {Marconi}, A. and {Netzer}, H. and {Shemmer}, O. and {Cresci}, G.},
        title = "{Observational evidence of quasar feedback quenching star formation at high redshift}",
      journal = {\aap},
         year = 2012,
        month = jan,
       volume = {537},
          eid = {L8},
        pages = {L8},
          doi = {10.1051/0004-6361/201118358},
archivePrefix = {arXiv},
       eprint = {1112.3071},
 primaryClass = {astro-ph.CO},
       adsurl = {https://ui.adsabs.harvard.edu/abs/2012A&A...537L...8C}
}

@ARTICLE{Nesvadba2011,
       author = {{Nesvadba}, N.~P.~H. and {Polletta}, M. and {Lehnert}, M.~D. and {Bergeron}, J. and {De Breuck}, C. and {Lagache}, G. and {Omont}, A.},
        title = "{The dynamics of the ionized and molecular interstellar medium in powerful obscured quasars at z{\ensuremath{\geq}} 3.5}",
      journal = {\mnras},
         year = 2011,
        month = aug,
       volume = {415},
       number = {3},
        pages = {2359-2372},
          doi = {10.1111/j.1365-2966.2011.18862.x},
archivePrefix = {arXiv},
       eprint = {1104.0937},
 primaryClass = {astro-ph.CO},
       adsurl = {https://ui.adsabs.harvard.edu/abs/2011MNRAS.415.2359N}
}

@ARTICLE{Alexander2010,
       author = {{Alexander}, D.~M. and {Swinbank}, A.~M. and {Smail}, Ian and {McDermid}, R. and {Nesvadba}, N.~P.~H.},
        title = "{Searching for evidence of energetic feedback in distant galaxies: a galaxy wide outflow in a z \raisebox{-0.5ex}\textasciitilde 2 ultraluminous infrared galaxy}",
      journal = {\mnras},
         year = 2010,
        month = mar,
       volume = {402},
       number = {4},
        pages = {2211-2220},
          doi = {10.1111/j.1365-2966.2009.16046.x},
archivePrefix = {arXiv},
       eprint = {0911.0014},
 primaryClass = {astro-ph.CO},
       adsurl = {https://ui.adsabs.harvard.edu/abs/2010MNRAS.402.2211A}
}

@ARTICLE{Prochaska2009,
       author = {{Prochaska}, J. Xavier and {Hennawi}, Joseph F.},
        title = "{Quasars Probing Quasars. III. New Clues to Feedback, Quenching, and the Physics of Massive Galaxy Formation}",
      journal = {\apj},
         year = 2009,
        month = jan,
       volume = {690},
       number = {2},
        pages = {1558-1584},
          doi = {10.1088/0004-637X/690/2/1558},
archivePrefix = {arXiv},
       eprint = {0806.0862},
 primaryClass = {astro-ph},
       adsurl = {https://ui.adsabs.harvard.edu/abs/2009ApJ...690.1558P}
}

@ARTICLE{Nesvadba2008,
       author = {{Nesvadba}, N.~P.~H. and {Lehnert}, M.~D. and {De Breuck}, C. and {Gilbert}, A.~M. and {van Breugel}, W.},
        title = "{Evidence for powerful AGN winds at high redshift: dynamics of galactic outflows in radio galaxies during the ``Quasar Era''}",
      journal = {\aap},
         year = 2008,
        month = nov,
       volume = {491},
       number = {2},
        pages = {407-424},
          doi = {10.1051/0004-6361:200810346},
archivePrefix = {arXiv},
       eprint = {0809.5171},
 primaryClass = {astro-ph},
       adsurl = {https://ui.adsabs.harvard.edu/abs/2008A&A...491..407N}
}

@ARTICLE{Gopal-Krishna2001,
       author = {{Gopal-Krishna} and {Wiita}, Paul J.},
        title = "{Was the Cosmic Web of Protogalactic Material Permeated by Lobes of Radio Galaxies During the Quasar Era?}",
      journal = {\apjl},
         year = 2001,
        month = oct,
       volume = {560},
       number = {2},
        pages = {L115-L118},
          doi = {10.1086/324310},
archivePrefix = {arXiv},
       eprint = {astro-ph/0108117},
 primaryClass = {astro-ph},
       adsurl = {https://ui.adsabs.harvard.edu/abs/2001ApJ...560L.115G}
}

@ARTICLE{Magain1988,
       author = {{Magain}, P. and {Surdej}, J. and {Swings}, J. -P. and {Borgeest}, U. and {Kayser}, R.},
        title = "{Discovery of a quadruply lensed quasar: the 'clover leaf H1413 + 117}",
      journal = {\nat},
         year = 1988,
        month = jul,
       volume = {334},
       number = {6180},
        pages = {325-327},
          doi = {10.1038/334325a0},
       adsurl = {https://ui.adsabs.harvard.edu/abs/1988Natur.334..325M}
}

@ARTICLE{Granato1996,
       author = {{Granato}, Gian Luigi and {Danese}, Luigi and {Franceschini}, Alberto},
        title = "{Dust-enshrouded AGN Models for Hyperluminous, High-Redshift Infrared Galaxies}",
      journal = {\apjl},
         year = 1996,
        month = mar,
       volume = {460},
        pages = {L11},
          doi = {10.1086/309977},
archivePrefix = {arXiv},
       eprint = {astro-ph/9601082},
 primaryClass = {astro-ph},
       adsurl = {https://ui.adsabs.harvard.edu/abs/1996ApJ...460L..11G}
}

@ARTICLE{Kayser1990,
       author = {{Kayser}, R. and {Surdej}, J. and {Condon}, J.~J. and {Kellermann}, K.~I. and {Magain}, P. and {Remy}, M. and {Smette}, A.},
        title = "{New Observations and Gravitational Lens Models of the Cloverleaf Quasar H1413+117}",
      journal = {\apj},
         year = 1990,
        month = nov,
       volume = {364},
        pages = {15},
          doi = {10.1086/169382},
       adsurl = {https://ui.adsabs.harvard.edu/abs/1990ApJ...364...15K}
}

@INPROCEEDINGS{Joye2003,
       author = {{Joye}, W.~A. and {Mandel}, E.},
        title = "{New Features of SAOImage DS9}",
    booktitle = {Astronomical Data Analysis Software and Systems XII},
         year = 2003,
       editor = {{Payne}, H.~E. and {Jedrzejewski}, R.~I. and {Hook}, R.~N.},
       series = {Astronomical Society of the Pacific Conference Series},
       volume = {295},
        month = jan,
        pages = {489},
       adsurl = {https://ui.adsabs.harvard.edu/abs/2003ASPC..295..489J}
}

@ARTICLE{Cutri2003,
       author = {{Cutri}, R.~M. and {Skrutskie}, M.~F. and {van Dyk}, S. and {Beichman}, C.~A. and {Carpenter}, J.~M. and {Chester}, T. and {Cambresy}, L. and {Evans}, T. and {Fowler}, J. and {Gizis}, J. and {Howard}, E. and {Huchra}, J. and {Jarrett}, T. and {Kopan}, E.~L. and {Kirkpatrick}, J.~D. and {Light}, R.~M. and {Marsh}, K.~A. and {McCallon}, H. and {Schneider}, S. and {Stiening}, R. and {Sykes}, M. and {Weinberg}, M. and {Wheaton}, W.~A. and {Wheelock}, S. and {Zacarias}, N.},
        title = "{VizieR Online Data Catalog: 2MASS All-Sky Catalog of Point Sources (Cutri+ 2003)}",
      journal = {VizieR Online Data Catalog},
         year = 2003,
        month = jun,
          eid = {II/246},
        pages = {II/246},
       adsurl = {https://ui.adsabs.harvard.edu/abs/2003yCat.2246....0C}
}

@ARTICLE{Lanzetta2002,
       author = {{Lanzetta}, Kenneth M. and {Yahata}, Noriaki and {Pascarelle}, Sebastian and {Chen}, Hsiao-Wen and {Fern{\'a}ndez-Soto}, Alberto},
        title = "{The Star Formation Rate Intensity Distribution Function: Implications for the Cosmic Star Formation Rate History of the Universe}",
      journal = {\apj},
         year = 2002,
        month = may,
       volume = {570},
       number = {2},
        pages = {492-501},
          doi = {10.1086/339774},
archivePrefix = {arXiv},
       eprint = {astro-ph/0111129},
 primaryClass = {astro-ph},
       adsurl = {https://ui.adsabs.harvard.edu/abs/2002ApJ...570..492L}
}

@ARTICLE{McNamara2007,
       author = {{McNamara}, B.~R. and {Nulsen}, P.~E.~J.},
        title = "{Heating Hot Atmospheres with Active Galactic Nuclei}",
      journal = {\araa},
         year = 2007,
        month = sep,
       volume = {45},
       number = {1},
        pages = {117-175},
          doi = {10.1146/annurev.astro.45.051806.110625},
archivePrefix = {arXiv},
       eprint = {0709.2152},
 primaryClass = {astro-ph},
       adsurl = {https://ui.adsabs.harvard.edu/abs/2007ARA&A..45..117M}
}

@ARTICLE{Springe2005,
       author = {{Springel}, Volker and {Di Matteo}, Tiziana and {Hernquist}, Lars},
        title = "{Modelling feedback from stars and black holes in galaxy mergers}",
      journal = {\mnras},
         year = 2005,
        month = aug,
       volume = {361},
       number = {3},
        pages = {776-794},
          doi = {10.1111/j.1365-2966.2005.09238.x},
archivePrefix = {arXiv},
       eprint = {astro-ph/0411108},
 primaryClass = {astro-ph},
       adsurl = {https://ui.adsabs.harvard.edu/abs/2005MNRAS.361..776S}
}

@ARTICLE{Cattaneo2009,
       author = {{Cattaneo}, A. and {Faber}, S.~M. and {Binney}, J. and {Dekel}, A. and {Kormendy}, J. and {Mushotzky}, R. and {Babul}, A. and {Best}, P.~N. and {Br{\"u}ggen}, M. and {Fabian}, A.~C. and {Frenk}, C.~S. and {Khalatyan}, A. and {Netzer}, H. and {Mahdavi}, A. and {Silk}, J. and {Steinmetz}, M. and {Wisotzki}, L.},
        title = "{The role of black holes in galaxy formation and evolution}",
      journal = {\nat},
         year = 2009,
        month = jul,
       volume = {460},
       number = {7252},
        pages = {213-219},
          doi = {10.1038/nature08135},
archivePrefix = {arXiv},
       eprint = {0907.1608},
 primaryClass = {astro-ph.CO},
       adsurl = {https://ui.adsabs.harvard.edu/abs/2009Natur.460..213C}
}

@ARTICLE{Shi2021,
       author = {{Shi}, Fangzheng and {Li}, Zhiyuan and {Yuan}, Feng and {Zhu}, Bocheng},
        title = "{An energetic hot wind from the low-luminosity active galactic nucleus M81*}",
      journal = {Nature Astronomy},
         year = 2021,
        month = jul,
       volume = {5},
        pages = {928-935},
          doi = {10.1038/s41550-021-01394-0},
archivePrefix = {arXiv},
       eprint = {2106.04041},
 primaryClass = {astro-ph.HE},
       adsurl = {https://ui.adsabs.harvard.edu/abs/2021NatAs...5..928S}
}

@ARTICLE{Cattaneo2009MN,
       author = {{Cattaneo}, A. and {Best}, P.~N.},
        title = "{On the jet contribution to the active galactic nuclei cosmic energy budget}",
      journal = {\mnras},
         year = 2009,
        month = may,
       volume = {395},
       number = {1},
        pages = {518-523},
          doi = {10.1111/j.1365-2966.2009.14557.x},
archivePrefix = {arXiv},
       eprint = {0812.1562},
 primaryClass = {astro-ph},
       adsurl = {https://ui.adsabs.harvard.edu/abs/2009MNRAS.395..518C}
}

@INPROCEEDINGS{Lawrence1996,
       author = {{Lawrence}, C.~R.},
        title = "{Observations Of Lens Systems With Keck I}",
    booktitle = {Astrophysical Applications of Gravitational Lensing},
         year = 1996,
       editor = {{Kochanek}, C.~S. and {Hewitt}, Jacqueline N.},
       volume = {173},
        month = jan,
        pages = {299},
       series = "{International Astronomical Union Symposia}",
       adsurl = {https://ui.adsabs.harvard.edu/abs/1996IAUS..173..299L}
}

@ARTICLE{Chantry2007,
       author = {{Chantry}, V. and {Magain}, P.},
        title = "{Deconvolution of HST images of the Cloverleaf gravitational lens. Detection of the lensing galaxy and a partial Einstein ring}",
      journal = {\aap},
         year = 2007,
        month = aug,
       volume = {470},
       number = {2},
        pages = {467-473},
          doi = {10.1051/0004-6361:20066839},
archivePrefix = {arXiv},
       eprint = {astro-ph/0612094},
 primaryClass = {astro-ph},
       adsurl = {https://ui.adsabs.harvard.edu/abs/2007A&A...470..467C}
}

@ARTICLE{Turnshek1988,
       author = {{Turnshek}, David A. and {Foltz}, Craig B. and {Grillmair}, Carl J. and {Weymann}, Ray J.},
        title = "{QSOs with PHL 5200--like Broad Absorption Line Profiles}",
      journal = {\apj},
         year = 1988,
        month = feb,
       volume = {325},
        pages = {651},
          doi = {10.1086/166036},
       adsurl = {https://ui.adsabs.harvard.edu/abs/1988ApJ...325..651T}
}

@ARTICLE{Angonin1990,
       author = {{Angonin}, M.~C. and {Remy}, M. and {Surdej}, J. and {Vanderriest}, C.},
        title = "{First spectroscopic evidence of microlensing on a BAL quasar ? The case of H 1413+117.}",
      journal = {\aap},
         year = 1990,
        month = jul,
       volume = {233},
        pages = {L5},
       adsurl = {https://ui.adsabs.harvard.edu/abs/1990A&A...233L...5A}
}

@ARTICLE{Turnshek1997,
       author = {{Turnshek}, David A. and {Lupie}, Olivia L. and {Rao}, Sandhya M. and {Espey}, Brian R. and {Sirola}, Christopher J.},
        title = "{Hubble Space Telescope Observations of the Gravitationally Lensed Cloverleaf Broad Absorption Line QSO H1413+1143: Imaging}",
      journal = {\apj},
         year = 1997,
        month = aug,
       volume = {485},
       number = {1},
        pages = {100-111},
          doi = {10.1086/304395},
       adsurl = {https://ui.adsabs.harvard.edu/abs/1997ApJ...485..100T}
}

@ARTICLE{Silverman2019,
       author = {{Silverman}, John D. and {Treu}, Tommaso and {Ding}, Xuheng and {Jahnke}, Knud and {Bennert}, Vardha N. and {Birrer}, Simon and {Schramm}, Malte and {Schulze}, Andreas and {Kartaltepe}, Jeyhan S. and {Sanders}, David B. and {Cen}, Renyue},
        title = "{Where Do Quasar Hosts Lie with Respect to the Size-Mass Relation of Galaxies?}",
      journal = {\apjl},
         year = 2019,
        month = dec,
       volume = {887},
       number = {1},
          eid = {L5},
        pages = {L5},
          doi = {10.3847/2041-8213/ab5851},
archivePrefix = {arXiv},
       eprint = {1910.14242},
 primaryClass = {astro-ph.GA},
       adsurl = {https://ui.adsabs.harvard.edu/abs/2019ApJ...887L...5S}
}

@ARTICLE{Li2021,
       author = {{Li}, Junyao and {Silverman}, John D. and {Ding}, Xuheng and {Strauss}, Michael A. and {Goulding}, Andy and {Birrer}, Simon and {Yesuf}, Hassen M. and {Xue}, Yongquan and {Kawinwanichakij}, Lalitwadee and {Matsuoka}, Yoshiki and {Toba}, Yoshiki and {Nagao}, Tohru and {Schramm}, Malte and {Inayoshi}, Kohei},
        title = "{The Sizes of Quasar Host Galaxies in the Hyper Suprime-Cam Subaru Strategic Program}",
      journal = {\apj},
         year = 2021,
        month = sep,
       volume = {918},
       number = {1},
          eid = {22},
        pages = {22},
          doi = {10.3847/1538-4357/ac06a8},
archivePrefix = {arXiv},
       eprint = {2105.06568},
 primaryClass = {astro-ph.GA},
       adsurl = {https://ui.adsabs.harvard.edu/abs/2021ApJ...918...22L}
}

@ARTICLE{Ding2022,
       author = {{Ding}, Xuheng and {Silverman}, John D. and {Onoue}, Masafusa},
        title = "{Opening the era of quasar host studies at high redshift with JWST}",
      journal = {arXiv e-prints},
         year = 2022,
        month = sep,
          eid = {arXiv:2209.03359},
        pages = {arXiv:2209.03359},
archivePrefix = {arXiv},
       eprint = {2209.03359},
 primaryClass = {astro-ph.GA},
       adsurl = {https://ui.adsabs.harvard.edu/abs/2022arXiv220903359D}
}

@ARTICLE{Stacey2022,
       author = {{Stacey}, H.~R. and {Arrigoni Battaia}, F.},
        title = "{Luck of the Irish? A companion of the Cloverleaf connected by a bridge of molecular gas}",
      journal = {\mnras},
         year = 2022,
        month = nov,
       volume = {517},
       number = {1},
        pages = {L11-L15},
          doi = {10.1093/mnrasl/slac102},
archivePrefix = {arXiv},
       eprint = {2209.00012},
 primaryClass = {astro-ph.GA},
       adsurl = {https://ui.adsabs.harvard.edu/abs/2022MNRAS.517L..11S}
}

@ARTICLE{Tessore2015,
       author = {{Tessore}, Nicolas and {Metcalf}, R. Benton},
        title = "{The elliptical power law profile lens}",
      journal = {\aap},
         year = 2015,
        month = aug,
       volume = {580},
          eid = {A79},
        pages = {A79},
          doi = {10.1051/0004-6361/201526773},
archivePrefix = {arXiv},
       eprint = {1507.01819},
 primaryClass = {astro-ph.CO},
       adsurl = {https://ui.adsabs.harvard.edu/abs/2015A&A...580A..79T}
}

@ARTICLE{Planck2020,
       author = {{Planck Collaboration} and {Aghanim}, N. and {Akrami}, Y. and {Ashdown}, M. and {Aumont}, J. and {Baccigalupi}, C. and {Ballardini}, M. and {Banday}, A.~J. and {Barreiro}, R.~B. and {Bartolo}, N. and {Basak}, S. and {Battye}, R. and {Benabed}, K. and {Bernard}, J. -P. and {Bersanelli}, M. and {Bielewicz}, P. and {Bock}, J.~J. and {Bond}, J.~R. and {Borrill}, J. and {Bouchet}, F.~R. and {Boulanger}, F. and {Bucher}, M. and {Burigana}, C. and {Butler}, R.~C. and {Calabrese}, E. and {Cardoso}, J. -F. and {Carron}, J. and {Challinor}, A. and {Chiang}, H.~C. and {Chluba}, J. and {Colombo}, L.~P.~L. and {Combet}, C. and {Contreras}, D. and {Crill}, B.~P. and {Cuttaia}, F. and {de Bernardis}, P. and {de Zotti}, G. and {Delabrouille}, J. and {Delouis}, J. -M. and {Di Valentino}, E. and {Diego}, J.~M. and {Dor{\'e}}, O. and {Douspis}, M. and {Ducout}, A. and {Dupac}, X. and {Dusini}, S. and {Efstathiou}, G. and {Elsner}, F. and {En{\ss}lin}, T.~A. and {Eriksen}, H.~K. and {Fantaye}, Y. and {Farhang}, M. and {Fergusson}, J. and {Fernandez-Cobos}, R. and {Finelli}, F. and {Forastieri}, F. and {Frailis}, M. and {Fraisse}, A.~A. and {Franceschi}, E. and {Frolov}, A. and {Galeotta}, S. and {Galli}, S. and {Ganga}, K. and {G{\'e}nova-Santos}, R.~T. and {Gerbino}, M. and {Ghosh}, T. and {Gonz{\'a}lez-Nuevo}, J. and {G{\'o}rski}, K.~M. and {Gratton}, S. and {Gruppuso}, A. and {Gudmundsson}, J.~E. and {Hamann}, J. and {Handley}, W. and {Hansen}, F.~K. and {Herranz}, D. and {Hildebrandt}, S.~R. and {Hivon}, E. and {Huang}, Z. and {Jaffe}, A.~H. and {Jones}, W.~C. and {Karakci}, A. and {Keih{\"a}nen}, E. and {Keskitalo}, R. and {Kiiveri}, K. and {Kim}, J. and {Kisner}, T.~S. and {Knox}, L. and {Krachmalnicoff}, N. and {Kunz}, M. and {Kurki-Suonio}, H. and {Lagache}, G. and {Lamarre}, J. -M. and {Lasenby}, A. and {Lattanzi}, M. and {Lawrence}, C.~R. and {Le Jeune}, M. and {Lemos}, P. and {Lesgourgues}, J. and {Levrier}, F. and {Lewis}, A. and {Liguori}, M. and {Lilje}, P.~B. and {Lilley}, M. and {Lindholm}, V. and {L{\'o}pez-Caniego}, M. and {Lubin}, P.~M. and {Ma}, Y. -Z. and {Mac{\'\i}as-P{\'e}rez}, J.~F. and {Maggio}, G. and {Maino}, D. and {Mandolesi}, N. and {Mangilli}, A. and {Marcos-Caballero}, A. and {Maris}, M. and {Martin}, P.~G. and {Martinelli}, M. and {Mart{\'\i}nez-Gonz{\'a}lez}, E. and {Matarrese}, S. and {Mauri}, N. and {McEwen}, J.~D. and {Meinhold}, P.~R. and {Melchiorri}, A. and {Mennella}, A. and {Migliaccio}, M. and {Millea}, M. and {Mitra}, S. and {Miville-Desch{\^e}nes}, M. -A. and {Molinari}, D. and {Montier}, L. and {Morgante}, G. and {Moss}, A. and {Natoli}, P. and {N{\o}rgaard-Nielsen}, H.~U. and {Pagano}, L. and {Paoletti}, D. and {Partridge}, B. and {Patanchon}, G. and {Peiris}, H.~V. and {Perrotta}, F. and {Pettorino}, V. and {Piacentini}, F. and {Polastri}, L. and {Polenta}, G. and {Puget}, J. -L. and {Rachen}, J.~P. and {Reinecke}, M. and {Remazeilles}, M. and {Renzi}, A. and {Rocha}, G. and {Rosset}, C. and {Roudier}, G. and {Rubi{\~n}o-Mart{\'\i}n}, J.~A. and {Ruiz-Granados}, B. and {Salvati}, L. and {Sandri}, M. and {Savelainen}, M. and {Scott}, D. and {Shellard}, E.~P.~S. and {Sirignano}, C. and {Sirri}, G. and {Spencer}, L.~D. and {Sunyaev}, R. and {Suur-Uski}, A. -S. and {Tauber}, J.~A. and {Tavagnacco}, D. and {Tenti}, M. and {Toffolatti}, L. and {Tomasi}, M. and {Trombetti}, T. and {Valenziano}, L. and {Valiviita}, J. and {Van Tent}, B. and {Vibert}, L. and {Vielva}, P. and {Villa}, F. and {Vittorio}, N. and {Wandelt}, B.~D. and {Wehus}, I.~K. and {White}, M. and {White}, S.~D.~M. and {Zacchei}, A. and {Zonca}, A.},
        title = "{Planck 2018 results. VI. Cosmological parameters}",
      journal = {\aap},
         year = 2020,
        month = sep,
       volume = {641},
          eid = {A6},
        pages = {A6},
          doi = {10.1051/0004-6361/201833910},
archivePrefix = {arXiv},
       eprint = {1807.06209},
 primaryClass = {astro-ph.CO},
       adsurl = {https://ui.adsabs.harvard.edu/abs/2020A&A...641A...6P}
}

@ARTICLE{Hazard1984,
       author = {{Hazard}, C. and {Morton}, D.~C. and {Terlevich}, R. and {McMahon}, R.},
        title = "{Nine new quasi-stellar objects with broad absorption lines.}",
      journal = {\apj},
         year = 1984,
        month = jul,
       volume = {282},
        pages = {33-52},
          doi = {10.1086/162174},
       adsurl = {https://ui.adsabs.harvard.edu/abs/1984ApJ...282...33H}
}

@ARTICLE{Chartas2004,
       author = {{Chartas}, G. and {Eracleous}, M. and {Agol}, E. and {Gallagher}, S.~C.},
        title = "{Chandra Observations of the Cloverleaf Quasar H1413+117: A Unique Laboratory for Microlensing Studies of a LoBAL Quasar}",
      journal = {\apj},
         year = 2004,
        month = may,
       volume = {606},
       number = {1},
        pages = {78-84},
          doi = {10.1086/382743},
archivePrefix = {arXiv},
       eprint = {astro-ph/0401240},
 primaryClass = {astro-ph},
       adsurl = {https://ui.adsabs.harvard.edu/abs/2004ApJ...606...78C}
}

@ARTICLE{Spingola2018,
       author = {{Spingola}, C. and {McKean}, J.~P. and {Auger}, M.~W. and {Fassnacht}, C.~D. and {Koopmans}, L.~V.~E. and {Lagattuta}, D.~J. and {Vegetti}, S.},
        title = "{SHARP - V. Modelling gravitationally lensed radio arcs imaged with global VLBI observations}",
      journal = {\mnras},
         year = 2018,
        month = aug,
       volume = {478},
       number = {4},
        pages = {4816-4829},
          doi = {10.1093/mnras/sty1326},
archivePrefix = {arXiv},
       eprint = {1807.05566},
 primaryClass = {astro-ph.GA},
       adsurl = {https://ui.adsabs.harvard.edu/abs/2018MNRAS.478.4816S}
}

@ARTICLE{Powell2022,
       author = {{Powell}, Devon M. and {Vegetti}, Simona and {McKean}, J.~P. and {Spingola}, Cristiana and {Stacey}, Hannah R. and {Fassnacht}, Christopher D.},
        title = "{A lensed radio jet at milliarcsecond resolution I: Bayesian comparison of parametric lens models}",
      journal = {\mnras},
         year = 2022,
        month = oct,
       volume = {516},
       number = {2},
        pages = {1808-1828},
          doi = {10.1093/mnras/stac2350},
archivePrefix = {arXiv},
       eprint = {2207.03375},
 primaryClass = {astro-ph.GA},
       adsurl = {https://ui.adsabs.harvard.edu/abs/2022MNRAS.516.1808P}
}

@ARTICLE{Condon1992,
       author = {{Condon}, J.~J.},
        title = "{Radio emission from normal galaxies.}",
      journal = {\araa},
         year = 1992,
        month = jan,
       volume = {30},
        pages = {575-611},
          doi = {10.1146/annurev.aa.30.090192.003043},
       adsurl = {https://ui.adsabs.harvard.edu/abs/1992ARA&A..30..575C}
}




\appendix

\section{Mock data} \label{App:Mock}
In Figure \ref{fig:caustics}, we present the mock data of the lensing system of
Cloverleaf, to simulate the response of a background source with a S\'{e}rsic
profile.  The lens model is the same as the one in our non-parametric model.
The source is a S\'{e}rsic ellipsoid moving from the northwest part to the
southeast part of the inner caustics in the source plane, and its centre is
marked with red crosses as moving.

\begin{figure*}
    \centering
    \includegraphics[scale=0.2]{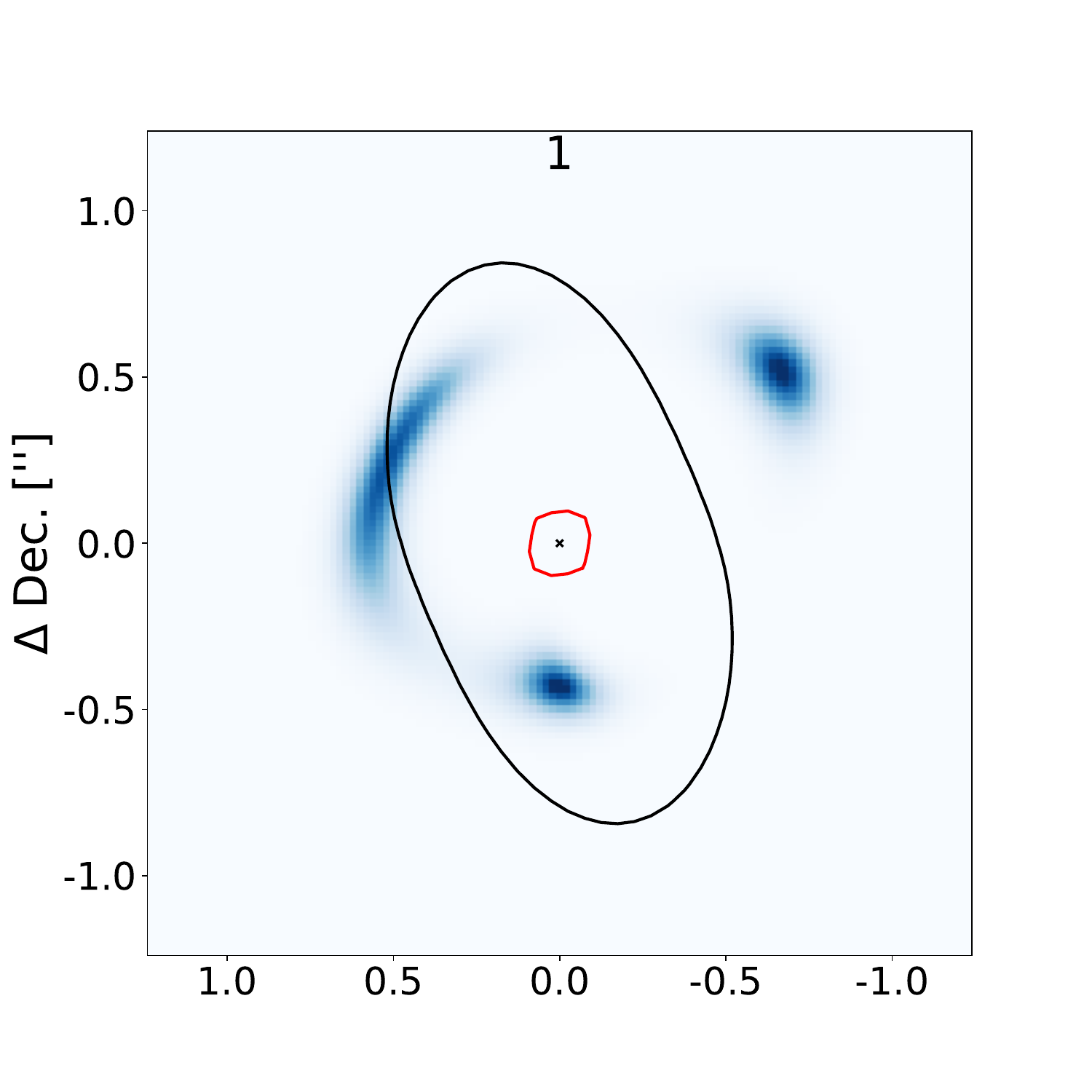}
    \includegraphics[scale=0.2]{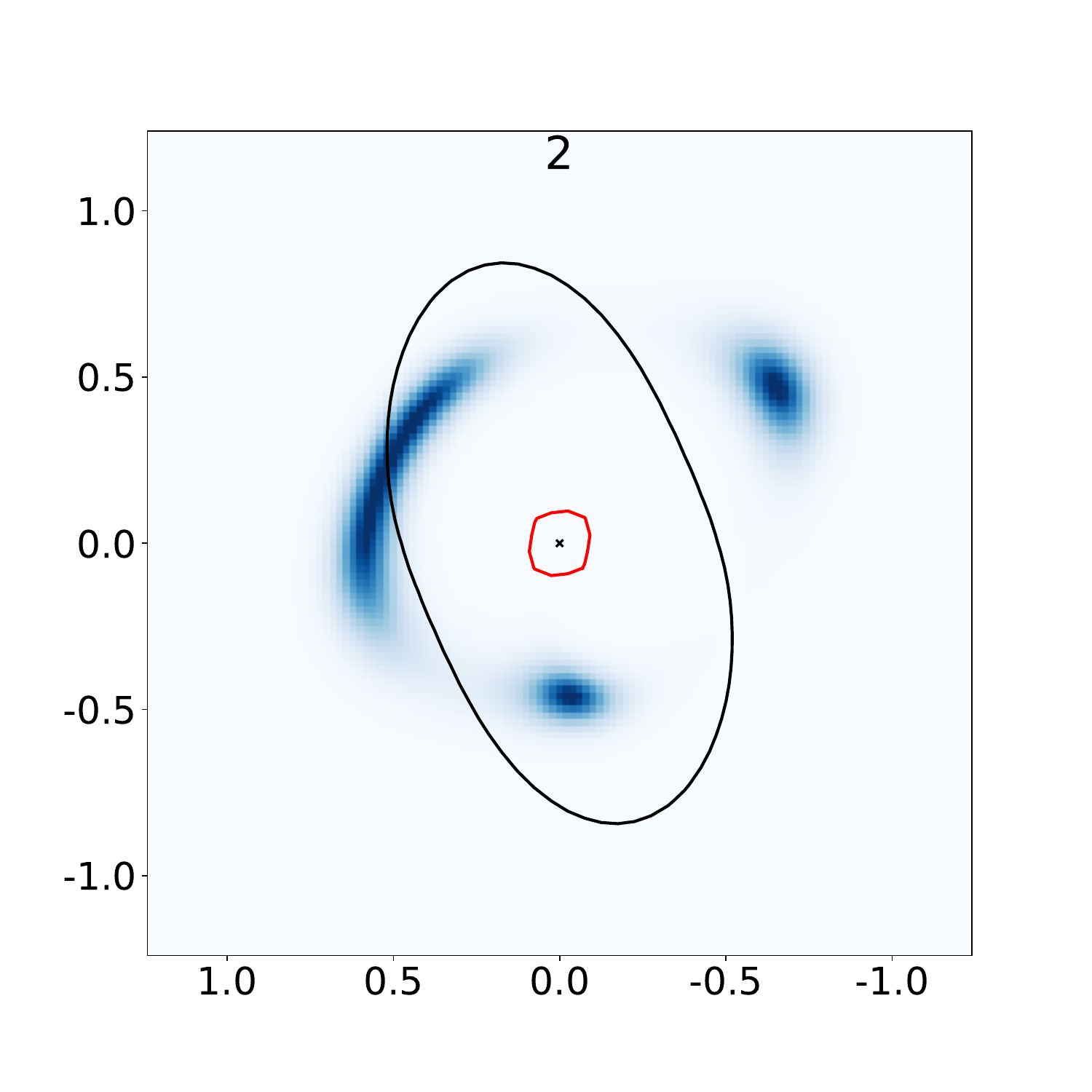}
    \includegraphics[scale=0.2]{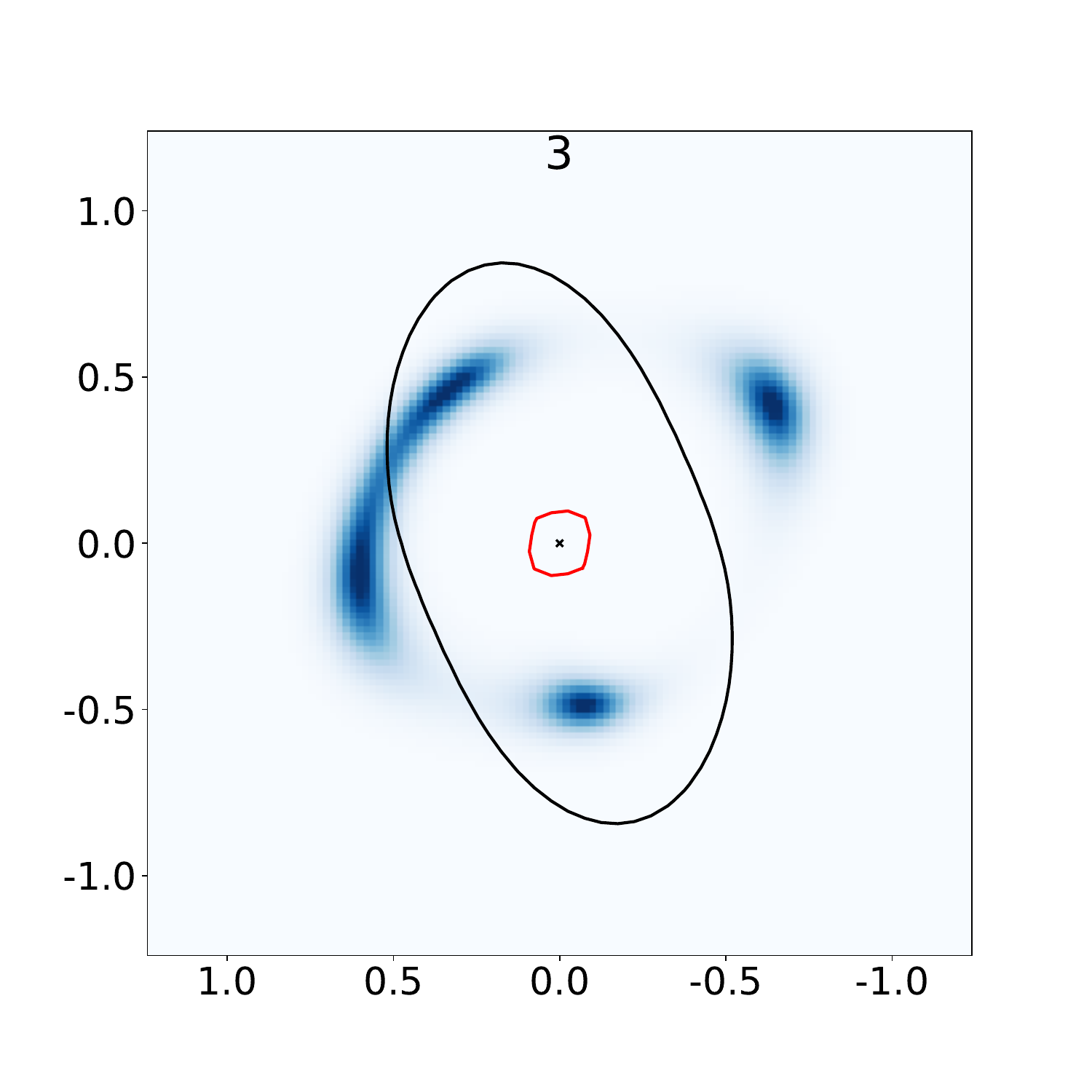}
    \includegraphics[scale=0.2]{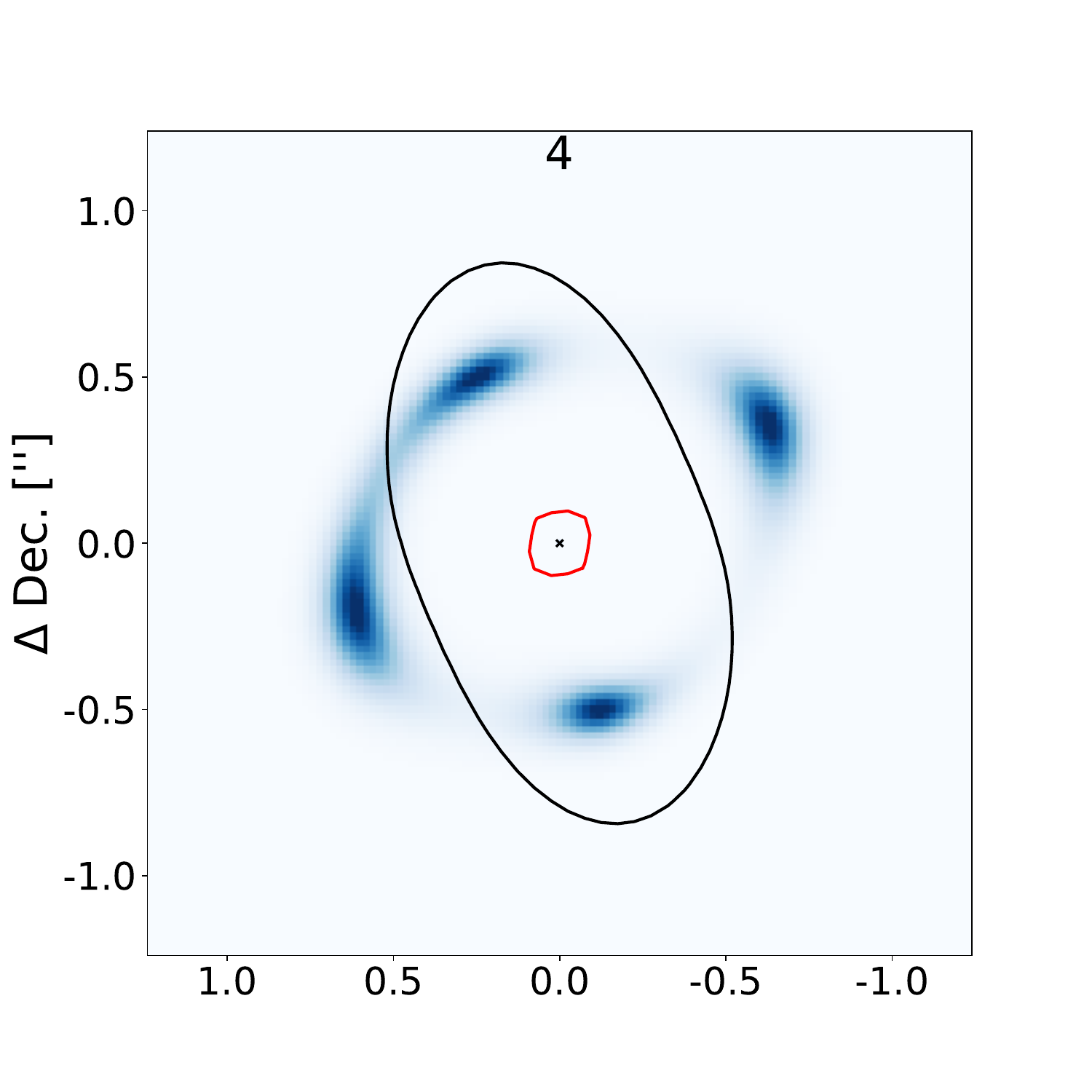}
    \includegraphics[scale=0.2]{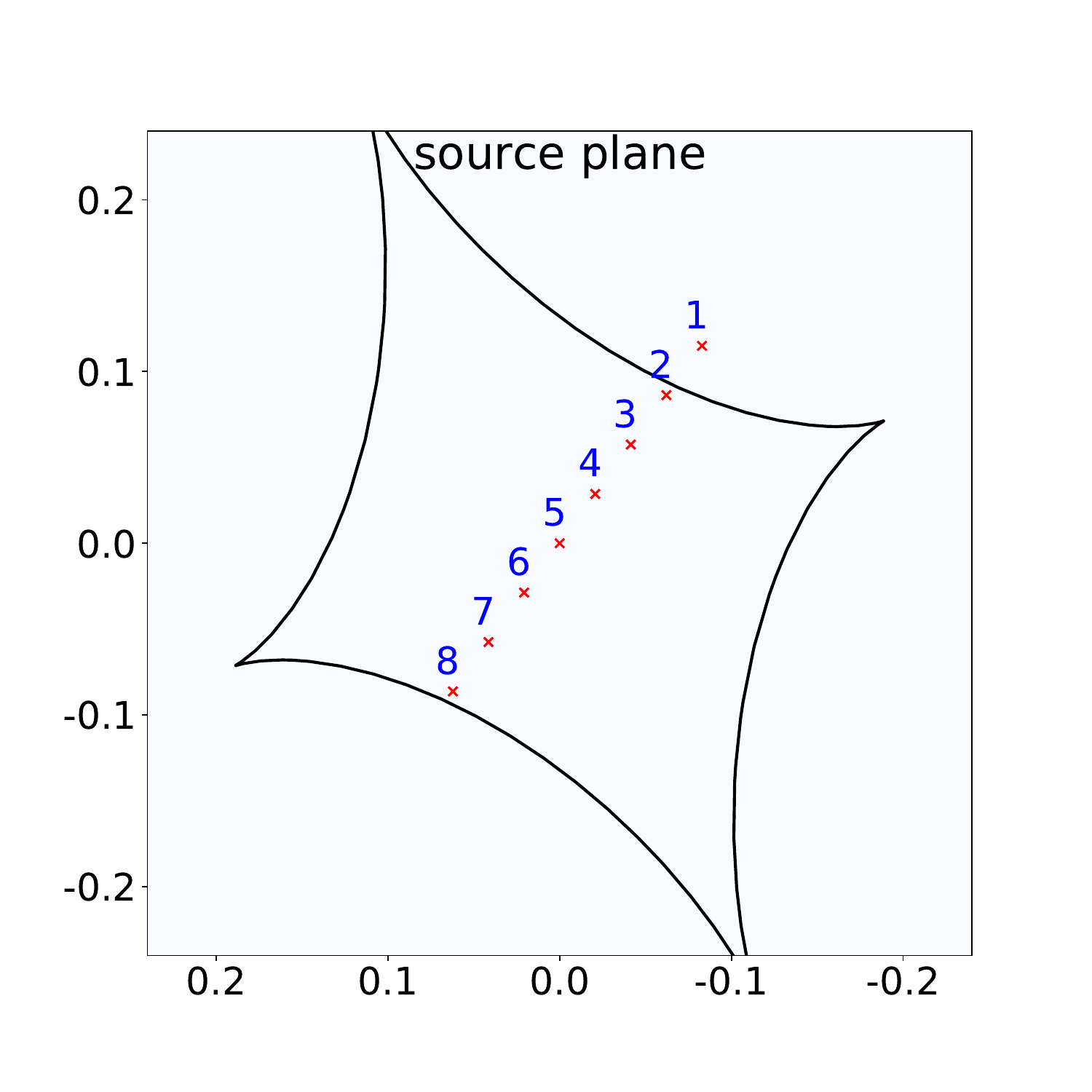}
    \includegraphics[scale=0.2]{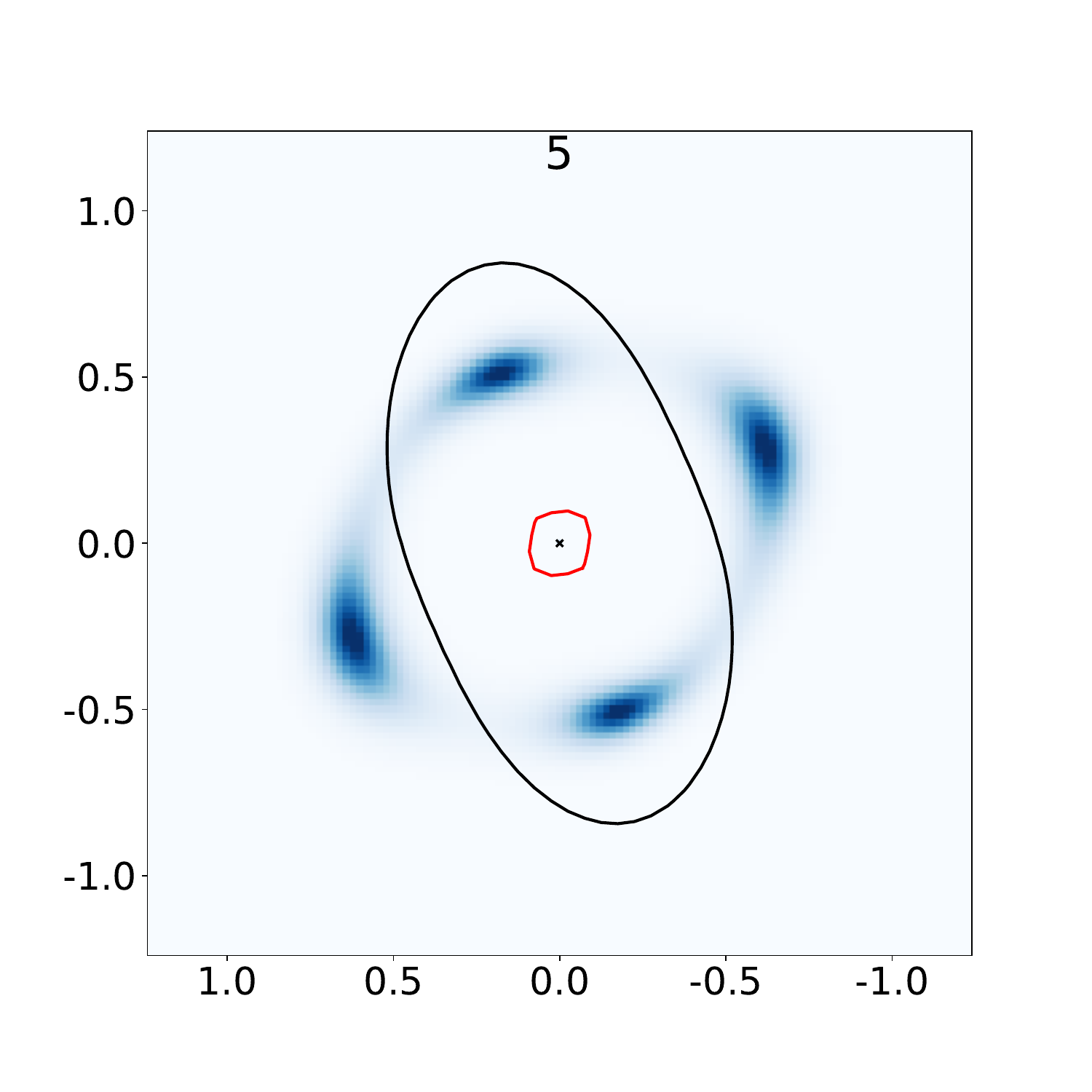}
    \includegraphics[scale=0.2]{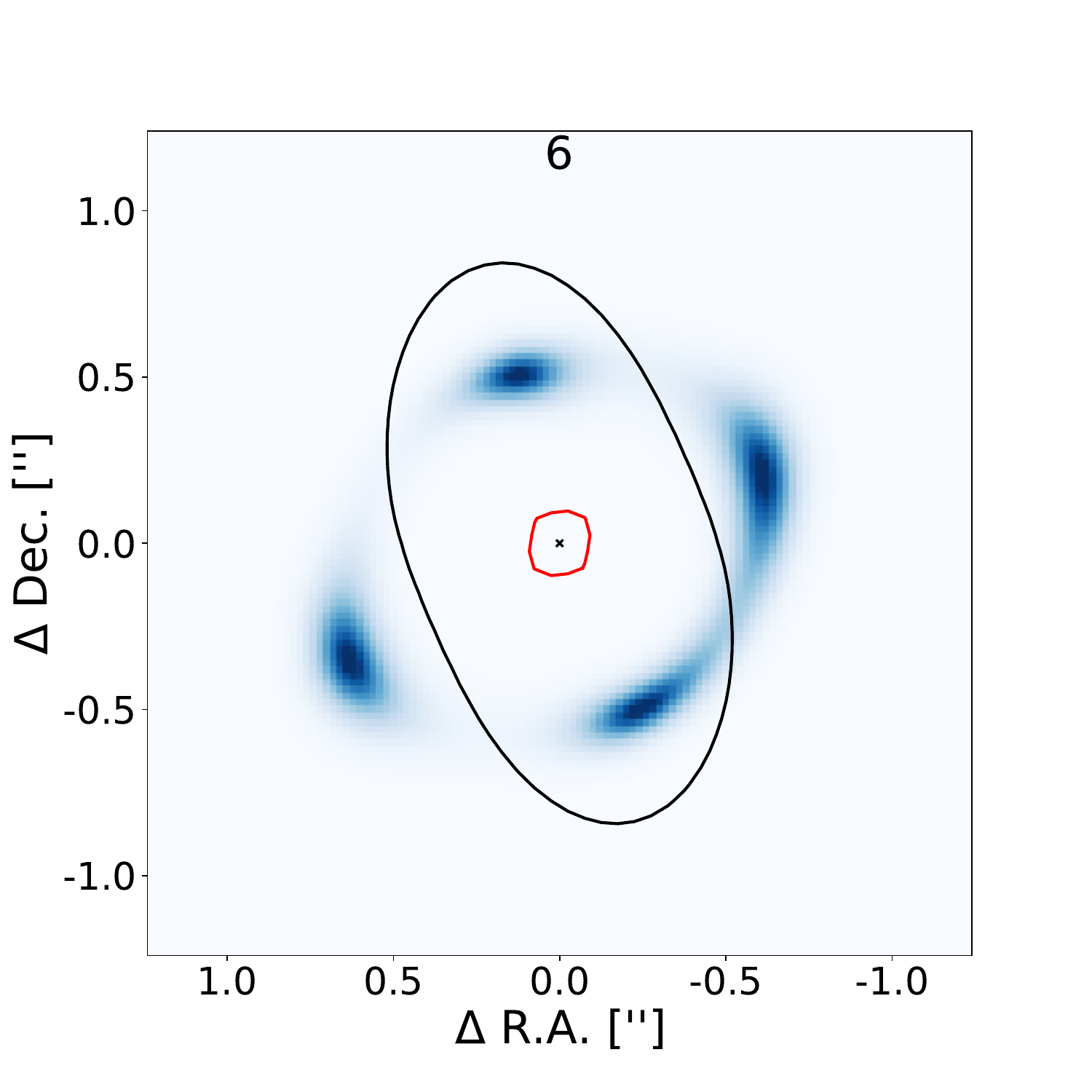}
    \includegraphics[scale=0.2]{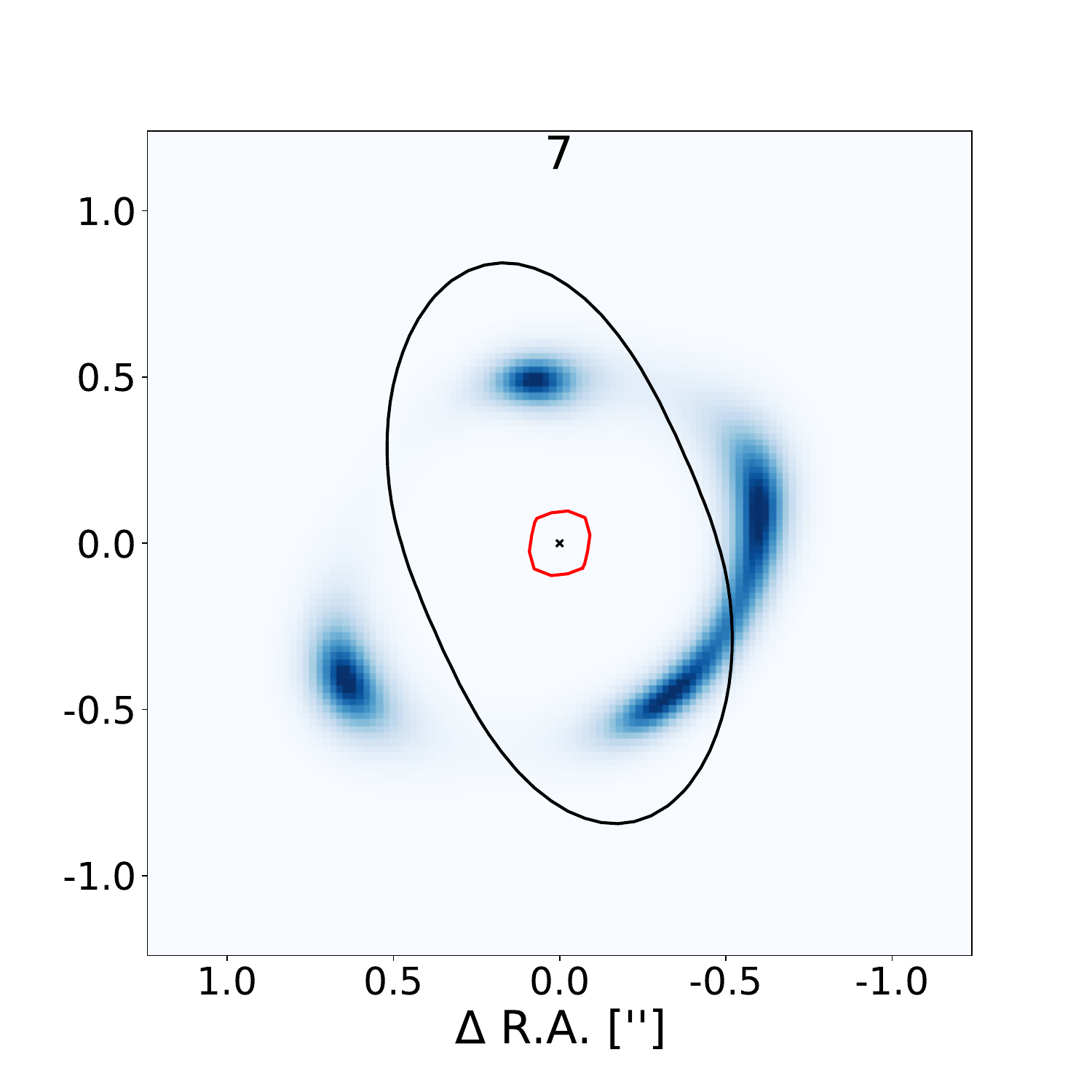}
    \includegraphics[scale=0.2]{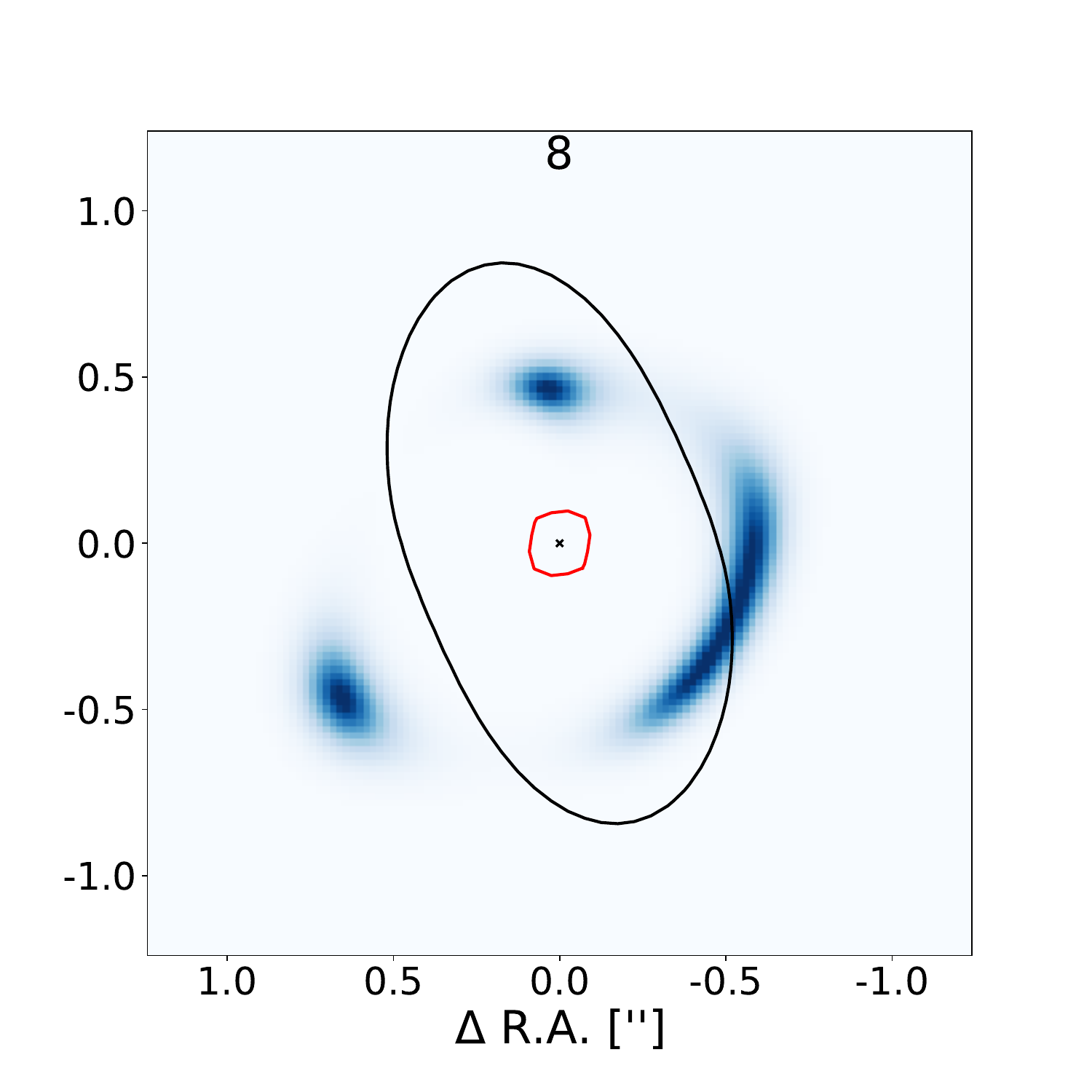}
    \caption{Mock data of this lensing system: The lens model is the same as
            the one in our non-parametric model. The source is a S\'{e}rsic
    ellipsoid moving from the northwest part to the southeast part of the
    inner caustics in the source plane, and its centre is marked with red
    crosses as moving. The solid black line and the red solid line are the
    outer and the inner critical curve (which is incorrect due to the numerical
    issue and should be a point) in the lens plane respectively. The solid black
    line (diamond-shaped) in the source plane is the inner caustics. }
    \label{fig:caustics}
\end{figure*}

\clearpage

\section{residual} \label{App:Residual}

In Figure \ref{fig:residuals}, we present the residual maps of our lensing
models of the  300 GHz, 1.5 GHz, 8.4 GHz, and 33 GHz data. 

\begin{figure*}
    \centering
    \includegraphics[scale=0.25]{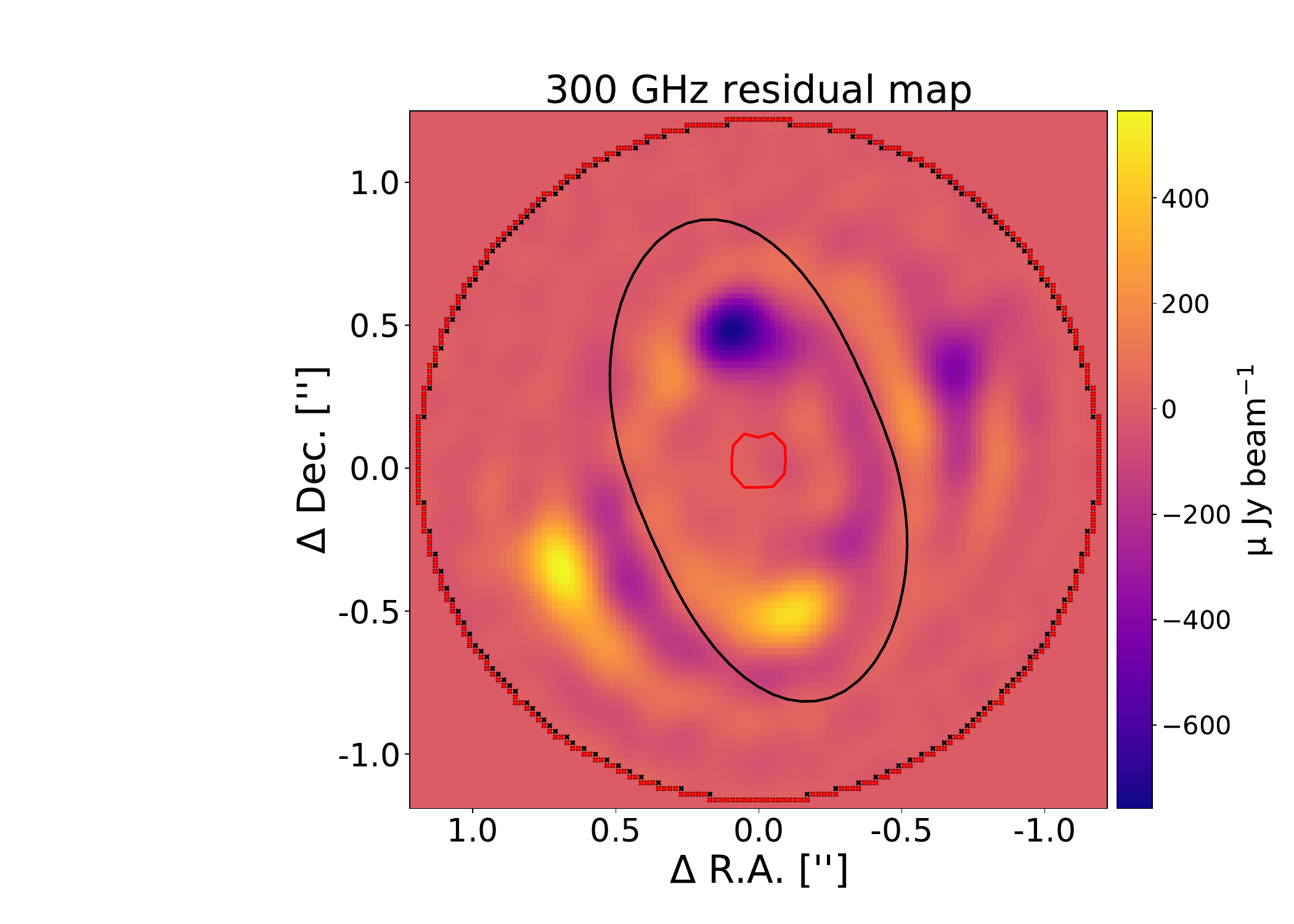}
    \includegraphics[scale=0.25]{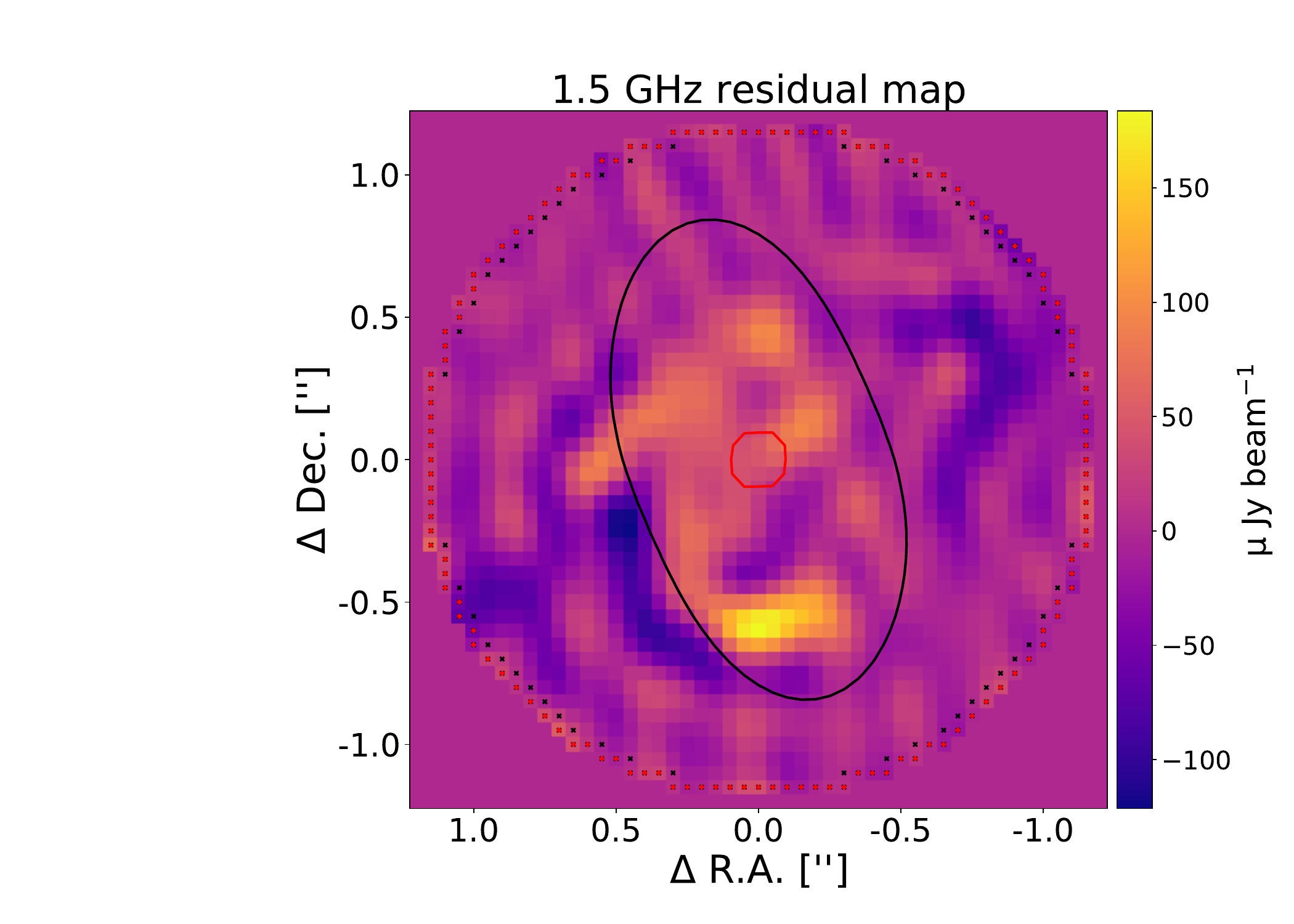}
    \includegraphics[scale=0.25]{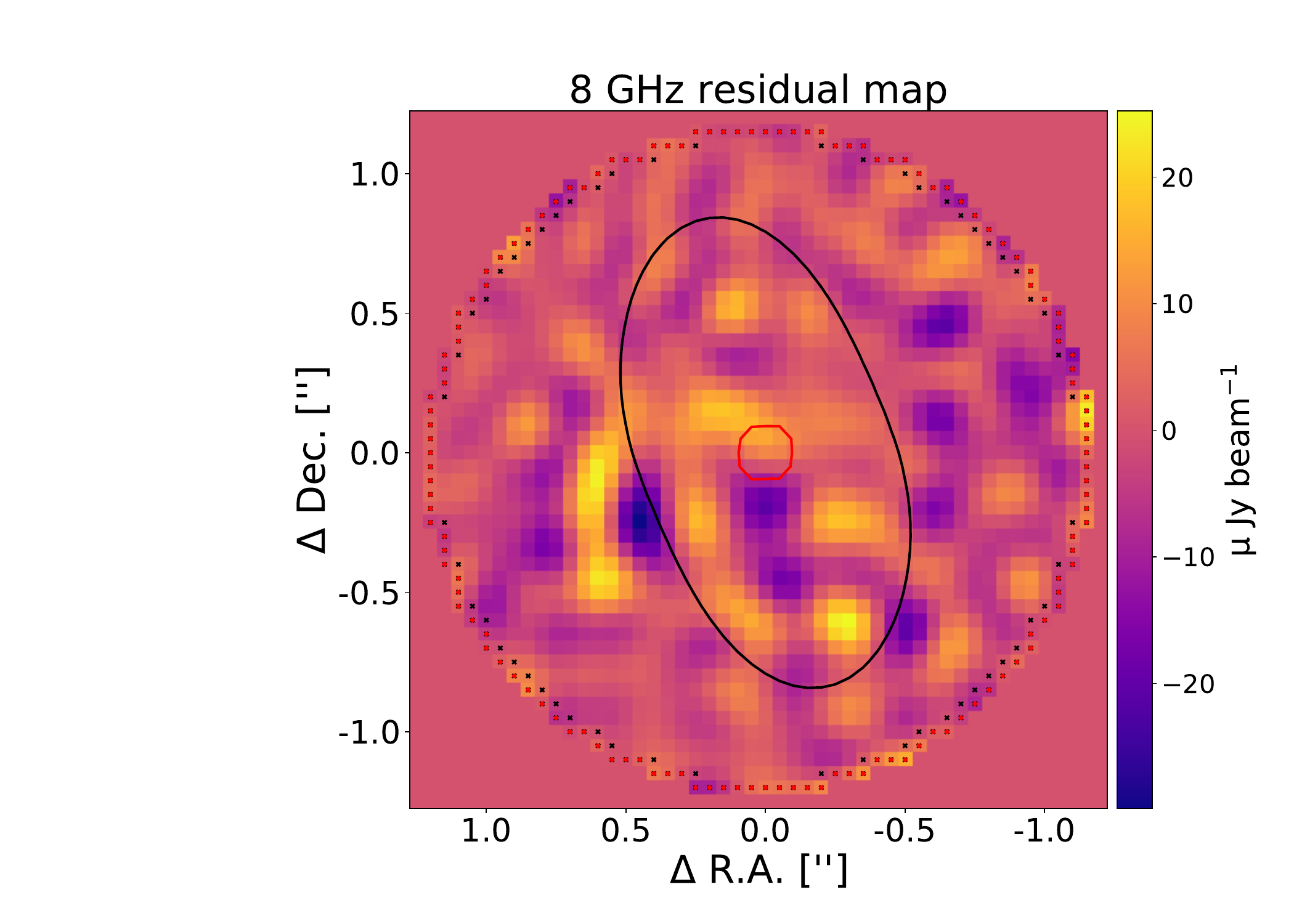}
    \includegraphics[scale=0.25]{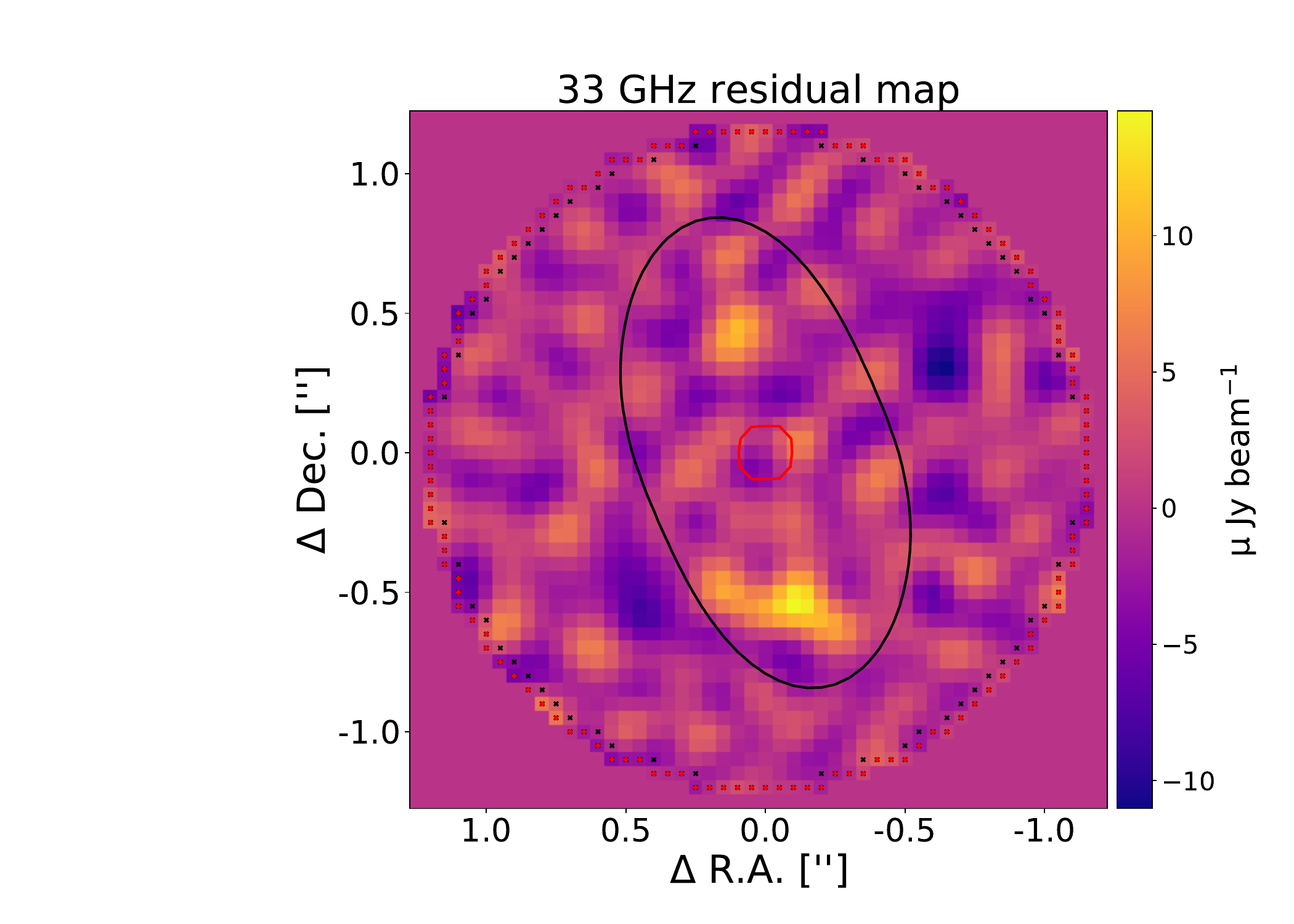}
    \caption{Residual maps of non-parametric models at 300 GHz, 1.5 GHz, 8.4 GHz,
    and 33 GHz: The red dot circle is the mask. The solid black line and the
    red solid line are the outer critical curve and the inner critical curve
    (which is incorrect due to the numerical issue and should be a point)
    respectively. } \label{fig:residuals}
\end{figure*}


\bsp	
\label{lastpage}
\end{document}